\documentstyle[aas2pp4]{article}

\received{10 July 1998} 
\accepted{15 September 1998} 

\slugcomment{Accepted to be published in The Astrophysical Journal}

\lefthead{Gonz\'alez Delgado et al.} \righthead{NGC7714}

\begin{document}

\title{Multiwavelength study of the starburst galaxy NGC 7714. I: Ultraviolet-Optical spectroscopy$^1$}

\author{Rosa M. Gonz\'alez Delgado} \affil{Instituto de Astrof\'\i sica
de Andaluc\'\i a, Apdo. 3004, 18080 Granada, Spain} \affil{Space Telescope Science Institute,
3700 San Martin Drive, Baltimore, MD 21218} \affil{Electronic mail:
rosa@iaa.es, gonzalez@stsci.edu}

\author{Mar\'\i a Luisa Garc\'\i a-Vargas} \affil{GTC-project. Instituto de
Astrof\'\i sica de Canarias, V\'\i a Lactea, S/N, 38200 La  Laguna, Tenerife,
Spain } \affil{Electronic mail: mgarcia@iac.es}

\author{Jeff Goldader} \affil{Space Telescope Science Institute, 3700 San Martin
Drive, Baltimore, MD 21218} \affil{Electronic mail: goldader@stsci.edu}

\author{Claus Leitherer} \affil{Space Telescope Science Institute, 3700 San
Martin Drive, Baltimore, MD 21218} \affil{Electronic mail: leitherer@stsci.edu}

\author{Anna Pasquali }
\affil{ST-ECF/ESO, Karl-Schwarzschild-Strasse 2, D-85748 Garching bei Muenchen, Germany}
\affil{Electronic mail: apasqual@eso.org}

$^1$ Based on observations with the NASA/ESA Hubble Space Telescope obtained  at
the Space Telescope Science Institute, which is operated by AURA, Inc.,  under
NASA contract NAS5-26555. \newpage

\begin{abstract} 

We have studied the physical conditions in the central 300 pc of the proto-typical starburst
galaxy NGC 7714. Our analysis is based on ultraviolet spectroscopy with the HST+GHRS and ground-based optical observations, and it covers also X-ray and radio data taken from the 
literature. The data are interpreted using evolutionary models optimized  for young starburst
regions. The massive stellar population is derived in a self-consistent way using 
the continuum and {\em stellar} absorption lines in the ultraviolet and the {\em nebular} 
emission line optical spectrum.

The central starburst has an age of about 4.5 Myr, with little evidence for
an age spread. Wolf-Rayet features at the ultraviolet indicates a stellar 
population of $\sim$ 2000 Wolf-Rayet stars. The overall properties of the
newly formed stars are quite similar to those derived, e.g., in 30~Doradus.
A standard Salpeter IMF is consistent with all observational constraints.
The nucleus of NGC~7714 has a bolometric luminosity of 0.5-1$\times$10$^{10}$ L$\odot$
and a mass of 5-10$\times$10$^6$ M$\odot$ (if low-mass stars have formed).  We find evidence for spatial structure within the central 300~pc
sampled. Therefore it is unlikely that the nucleus of NGC~7714 hosts a single
 star cluster exceeding the properties of other known clusters.
Contrary to previous suggestions, we find no evidence for a nuclear supernova 
rate that would significantly exceed the total disk-integrated rate. About one  
supernova event per century is predicted. Most of them are associated with
the core-collapse of a hydrogen-free or -poor progenitor.
An older stellar generation, with ages of tens of Myr and older, is
suggested as well. This population is less concentrated towards the nucleus and
extends over kpc scales

\end{abstract} 


\section{Introduction: NGC 7714 a test case for evolutionary synthesis models}

Starburst galaxies are objects in which the total energetics are dominated by star formation and 
associated phenomena (Weedman 1983; Heckman 1998). This definition covers galaxies with a very wide
 range of properties, from blue compact dwarfs to ultraluminous IRAS starbursts. Typical masses 
(bolometric luminosities) range from 10$^6$ to 10$^{10}$ M$\odot$ (10$^9$ to 10$^{14}$ L$\odot$), 
corresponding the lowest limit to the mass of super-star-clusters and the highest limit to the 
mass of the infrared-luminous galaxies (Leitherer 1996).  Nearby starburst 
galaxies are ideal laboratories in which to explore fundamental questions about the local and 
global processes and effects of star formation. For example, is the initial mass function the
same in starburst galaxies as in quiescent star formation in the Milky Way? Is there a connection 
between starburst activity and active galactic nuclei? What can starbursts teach us about the process
by which the first galaxies formed?   

Starburst are powered by massive stars. Massive stars (M$\geq10$ M$\odot$) emit photons with 
energies of tens of eV which are 
absorbed and re-emitted in their stellar winds, producing ultraviolet resonance transitions. 
However, the stellar wind is optically thin to most of
the  ultraviolet  photons, that can travel tens of pc from the star before
they are  absorbed and photoionize the surrounding interstellar medium. Then, this
ionized gas cools  down via an emission line spectrum.
This is, in essence, the spectral
dichotomy picture of a starburst galaxy: a nebular emission line spectrum at optical
wavelengths and an absorption-line spectrum at wavelengths shorter than the
Balmer jump (Leitherer 1997). 
However, the presence of a starburst affects the entire spectrum and not only the UV-optical spectrum. Thus, supernova
remnants following the explosion of massive stars generate X-ray and radio continuum emission; red supergiant stars emit 
strongly in the near-IR continuum; the mid-IR is dominated by line and continuum emission from warm dust within the star forming regions; the far-IR is thermal emission from dust heated by absorbing UV and optical light; electron accelerated by supernova remnants create radio synchroton emission. A truly complete analysis of a starburst would account for all these features.

Evolutionary synthesis is a powerful tool for understanding starburst galaxies. This  technique makes a prediction for the spectrum of a
stellar population taking as a free  parameter the star formation history of the
starburst (age, initial mass function, star formation rate, etc). If there is agreement between  
the observations
and the predicted physical properties (spectral energy distribution,  colors, line profiles, etc), 
constraints on the star formation history (in timescale between 1 Myr to several Gyr)
 can be determined (see e.g. Mas-Hesse \& Kunth, 1991;
Bruzual \& Charlot, 1993; Leitherer \& Heckman, 1995). However, the 
solution obtained with this technique may not be unique, and only {\em consistency}
between models and observations can be derived. 

To make a complete analysis of a starburst, we have to answer the question: are evolutionary
synthesis models, and  the stellar evolution models they incorporate, actually up to the
challenge of interpreting multi-wavelength data? To answer this, we have undertaken a detailed
study of the starburst galaxy NGC 7714. 

NGC 7714 is an SBb peculiar galaxy (de Vaucouleurs et al 1991) classified by Weedman et al 
(1981; here after W81) as a prototypical starburst galaxy. Its heliocentric radial velocity, as 
compiled by NASA Extragalactic Database (NED), is 2798 km s$^{-1}$. This places NGC 7714 at
a distance of 37.3 Mpc assuming H$_o$= 75 km s$^{-1}$ Mpc$^{-1}$. Heckman et al (1998) derive a
distance of 37.9 Mpc using the standard linear Virgo centric infall model of Schechter (1980)
with parameters $\gamma$=2, $v_{Virgo}$= 976 km s$^{-1}$, $w_{\odot}$=220 km s$^{-1}$ and
$D_{Virgo}$= 15.9 Mpc. The B magnitud and far-infrared luminosity (as calculated from the
60 $\mu$m and 100 $\mu$m IRAS bands) are -20.04 and 43.93 erg s$^{-1}$, respectively.  The
far-infrared ratio 60/100 and 25/60 are larger than 0.9 and 0.25, respectively.  The mass of
the  neutral gas is $\sim$ 10$^{10}$ M$\odot$ (Mirabel \& Wilson, 1989).  The X-ray, IUE
ultraviolet, optical, and VLA radio fluxes are explained as a consequence of intense star
formation activity in its nucleus. To explain the X-ray and radio luminosities, 10$^4$ supernova
remnants are required (W81). Bernl\"{o}hr (1993) found from the optical continuum and FIR colors
that the star formation in NGC 7714 is consistent with a continuous star formation rate during
the past 20 Myr, and that it had probably been triggered by the interaction with the companion
galaxy NGC 7715 (Vorontsov-Velyaminov 1977; Arp 1966; Smith, Struck, \& Pogge, 1997).  Very
detailed studies have been carried out in the optical and near-IR part of the spectrum to
characterize the gas properties in the nuclear and circumnuclear regions (Gonz\'alez-Delgado et
al 1995, hereafter GD95, and references therein). GD95 conclude that the burst of star formation
in the nucleus of the galaxy should have an age around 4-5 Myr, in view of the WR bump
detection, and pointed out the presence of a previous burst of star formation on the basis of
the detection of calcium triplet  absorption at 8600 \AA. Models by Garc\'{\i}a-Vargas et al
(1997) indicate that a young burst of age between 3.5 and 5 Myr is able to explain the emission
line spectrum and the WR luminosity bump in three circumnuclear HII regions of NGC 7714.  

Here, we present HST+GHRS ultraviolet spectroscopy of the nuclear starburst. These data are 
analyzed together with the optical emission line spectrum to obtain information about the 
ionizing stellar population. The goal is to perform a consistency test between the stellar
content derived from multi-wavelength observations in the central few hundred parsecs in NGC
7714. We will evaluate whether the youngest burst can account for the high supernova rate (1
yr$^{-1}$; W81) derived from the X-ray and radio continuum emission. A following paper will 
present the results for the NIR K-band spectroscopy and the analysis of the multiwavelength
spectral energy distribution (SED) (Goldader et al 1998; in preparation). It will focus on the
extincted and low mass stellar content of the nuclear starburst in NGC 7714.  Calzetti (1997) 
analyzed the properties of the NGC 7714 starburst on a kpc scale. The new GHRS data allow us
to probe the nuclear starburst on a scale that is an order of magnitud smaller. The new data
are useful to study if and how the star formation properties in NGC 7714 differ between pc and
kpc scales.

The paper is organized  as follows. Section 2 presents the  HST observations. Section 3 deals
with the results and interpretation of the HST ultraviolet spectrum. In Section 4 we present
photoionization models that fit the emission line spectrum in the optical-near infrared spectral
range. The multiwavelength analysis of the NGC 7714 spectrum is in Section 5. The summary and
the conclusions are in Section 6.  

\section{HST Observations} 

The ultraviolet spectrum of the central nuclear starburst in NGC 7714 was
obtained in October 1996 with the Hubble Space Telescope using the GHRS and the
grating G140L, through the Large Science Aperture (LSA, $1.74\times1.74$
arcsec$^2$ corresponding to 330$\times$330 pc$^{2}$). Two different wavelength settings were observed, covering 1175 to
1461 \AA\ and 1402 to 1689 \AA, with a total integration time of 1659 s and
4542 s, respectively. After standard pipeline processing, the spectra were
combined into a single spectrum covering 1175 to 1689 \AA, and rebinned to 0.57
\AA/pixel, which is the nominal dispersion of the grating. The spectrum was
corrected for redshift assuming a radial velocity of 2800 km s$^{-1}$ (GD95). Figure 1
shows the spectrum in the restframe wavelength system of the galaxy. 

The spectrum is rich in stellar (CIV $\lambda$1550, SV
$\lambda$1501, CIII $\lambda$1417, SiIV $\lambda$1400, NV $\lambda$1240, CIII
$\lambda$1175), and interstellar (AlII $\lambda$1670, FeII $\lambda$1608, CIV
$\lambda$1550, SiII $\lambda$1527, SiIV $\lambda$1400, CII $\lambda$1335,
OI+SiII $\lambda$1302, SiII $\lambda$1260) absorption lines. It also shows L$\alpha$ and HeII
$\lambda$1640 in emission. 

Some of the interstellar lines have weaker satellites formed in the Milky Way halo (SiII
$\lambda$1527, SiIV $\lambda$1400, CII $\lambda$1335, OI+SiII $\lambda$1302, and
SiII $\lambda$1260). The zero point of the wavelength calibration was checked by
fitting gaussians to the Galactic absorption lines. These lines are blueshifted
about 0.3-0.4 \AA\ with respect to the nominal zero point of --2800 km s$^{-1}$. This is well 
within the uncertainties expected for an observation through the LSA. The 
instrumental
spectral resolution for a point source within the LSA is about 0.8 \AA, but for an
extended source it depends on the size of the ultraviolet source. The widths (FWHM) of the
gaussians fitted to the Galactic lines are 3.3 \AA\ or 740 km s$^{-1}$ at 1400 \AA. This suggests 
that the object fills a substantial fraction of the aperture.

The flux measured at 1500 \AA\ is $1.9\times10^{-14}$ erg s$^{-1}$ cm$^{-2}$
\AA$^{-1}$. This 
is a factor of 2 lower than the IUE flux reported by Kinney et al
(1993), indicating that the UV bright stellar population significantly exceeds the area of the GHRS LSA.
An HST WFPC2 PC image through the filter F606W ($\rm\lambda_o$= 6010.6 \AA, 
$\Delta\lambda$= 1497 \AA) was retrieved from the HST archive. The band pass of
this filter includes strong emission lines such as [OIII]$\lambda$5007 and H$\alpha$, 
so that the light
detected through this filter is not only continuum emission. The integration time
was 500 s. The sampling is 0.046 arcsec/pixel, corresponding to 8.7 pc/pixel. Figure 2a shows 
a view of the whole galaxy. Most of the emission comes from the nucleus with some contribution from the two spiral arms. This central emission is resolved into a compact knot surrounded by
several smaller knots. The GHRS spectroscopy suggests that the
starburst is extended at ultraviolet wavelengths as well. Therefore the spatial morphology in 
the optical and the  ultraviolet is rather similar.

\section{Stellar content in the nuclear starburst in NGC7714: fitting the HST UV
spectrum} 

\subsection{Wind absorption lines} 

Massive hot stars develop strong stellar winds due to radiation pressure in
ultraviolet resonance lines (Morton 1967; Lucy \& Solomon 1970). Typical wind
velocities in O stars are about 2000 km s$^{-1}$ to 3000 km s$^{-1}$
(Groenewegen, Lamers, \& Pauldrach 1989; Prinja, Barlow, \& Howarth 1990). As a result, all
strong ultraviolet lines in the spectra of O stars originate predominantly in
the outflow, and have blueshifted absorptions. The profile shapes reflect the
stellar mass-loss rates, which are a strong function of the stellar luminosity
(Castor, Abbott, \& Klein 1975). Since there exists a well-defined stellar
mass-luminosity relation, the line profiles ultimately depend on the stellar
mass, and --- for a stellar population --- on the IMF and SF history (Leitherer, Robert, \& Heckman 1995). 

The strongest stellar wind resonance features are OVI $\lambda$1034, NV
$\lambda$1240, SiIV $\lambda$1400 and  CIV $\lambda$1550. In massive stars, these lines form above the
photosphere in the stellar wind --- either as a blueshifted absorption in stars
with weak winds, or as a P Cygni profile if the wind density is sufficiently
high. OVI and CIV are strong lines in O stars of all luminosity classes
(Walborn, Bohlin, \& Panek 1985; Gonz\'alez Delgado, Leitherer, \& Heckman
1997). In contrast, SiIV is luminosity dependent, and only  blue supergiant
stars produce a strong P Cygni profile (Walborn et al 1985). The recombination  line HeII $\lambda$1640 can also be formed in very massive O and WR stars with 
very dense winds. Evolutionary synthesis models for CIV (Leitherer, Robert, \& Heckman 1995) show that the line strength increases for a shallower IMF. The most
massive stars are the dominant contributors to the lines, and increasing their
numbers relative to the less massive stars which produce the continuum --- 
either via flattening the IMF slope or increasing the upper cut-off
mass --- leads to
stronger lines. SiIV shows a critical dependence on the age of the burst. This
spectroscopic method is only efficient in regions of recent star formation
activity and high stellar density where the highest masses can be sampled. The
technique has been succesfully applied to starburst galaxies (Conti, Leitherer, \& Vacca 1996; 
Leitherer et al 1996; Gonz\'alez Delgado et al 1998a) and Seyfert
galaxies (Heckman et al 1997; Gonz\'alez Delgado et al 1998b). In all these
cases, the method has constrained the age and the mass spectrum of the young
population. 

NGC 7714 has been observed previously by IUE (W81; Kinney et al 1993). The spectra show strong CIV and SiIV in absorption, indicating the presence of massive stars in NGC 7714. However, the interstellar lines are also very strong in NGC 7714 and contribute significantly to the strength of the CIV and SiIV lines. The interstellar contribution can some times dominate the profile of the line, as is the case of NGC 1705 (Heckman \& Leitherer 1997).  This is why an analysis based only in the equivalent width of these lines cannot predict  correctly the stellar content and the star formation history of the starburst. To fully resolve the stellar from the interstellar 
contribution, a resolution better than the IUE resolution of R=1000 is required.  

The most conspicuous stellar lines in the GHRS spectra of  NGC 7714 are the wind lines HeII, CIV, 
SiIV and NV. We use line profile synthesis of the ultraviolet wind lines to
derive the stellar content and to constrain the star formation history of the
nuclear starburst. We do not fit the entire line profile of the CIV and SiIV due
to the contribution of the interstellar lines, which are not fully accounted for in the
 models, and are stronger in the
starburst spectra than in the synthetic models. We select two windows,
corresponding to the blue side of the wind profile (1381.75-1390.75 \AA\  for
SiIV  and 1528.0-1543.75 \AA\ for CIV) and to the red side (1403.5-1409.5 \AA\ for
SiIV and 1552.0-1560.25 \AA\ for CIV), where the profile is dominated by the
stellar contribution. Then, we compute the $\chi^2$ parameter between the
observations and models. The results for SiIV are plotted for burst  and continuous star 
formation (csf) models as a function of the age, upper mass cut-off, and
slope of the IMF in Figures 3a,b  and 3c,d, respectively. The minimum $\chi^2$ values for
SiIV are obtained for instantaneous burst models and a Salpeter IMF. The fit for SiIV requires the presence of O supergiant stars, which appear between
3 and 6 Myr after the burst onset. The best fit is obtained for a 5 Myr burst. The
most massive stars present in the starburst have at least a mass of 40 M$\odot$. However,
we cannot constrain the upper mass cut-off because stars more
massive than 40 M$\odot$ have already reached the supernova phase at 5 Myr. Thus, the upper mass cut-off is higher than 40 M$\odot$. Figure 4
shows the profiles of the observed CIV and SiIV and the 5 Myr burst model
(Salpeter IMF and M$_{up}$=100 M$\odot$).  The CIV profile is equally well
fitted by csf models, but it is not  possible to constrain the age with these
models. 

Note that the emission of the NV P Cygni profile is weaker than predicted
by the 5 Myr burst  model. This discrepancy could be due to the SiII
$\lambda$1260 absorption line formed in the Milky Way halo that falls on the blue edge
of the emission part of the profile. 

The determination of the age is also consistent with the presence of WR stars in
the starburst. WR features have been detected at optical wavelengths (van Breugel et al 1985; GD95), and
also in our ultraviolet spectrum. The HeII $\lambda$1640 is a recombination line which
shows a broad emission profile if it is formed in the very dense stellar winds of
WR stars and O3-O5 supergiants (Walborn, Bohlin, \& Panek 1985). This line is prominent
in the GHRS spectrum, and the observed flux is $4\times10^{-14}$ erg s$^{-1}$
cm$^{-2}$. Later we will use this flux to estimate the number of WR stars 
present in the nuclear starburst.

\subsection{Photospheric absorption lines} 

The spectrum shows several photospheric lines, such as SV $\lambda$1501, CIII
$\lambda$1427, and SiIII $\lambda$1417. These lines are prominent in O and early
B stars (Walborn et al 1985; Walborn, Parker, \& Nichols 1995); they are not
resonance lines, and therefore they do not form in the interstellar medium.
Their measured equivalent widths are in Table 1. Note that the  strength of
these lines is also consistent with the synthetic model (see Figure 5).

\subsection{Extinction Estimates} 

To derive the number of WR and O stars in the starburst, we need to estimate the
extinction. This can be done using the ultraviolet continuum flux distribution. Leitherer
\& Heckman (1995) have derived from evolutionary synthesis models that the UV
energy distribution arising from a starburst has a spectral index, $\alpha$, in
the range -2.6 to -2.2 (F$_\lambda\propto$ $\lambda^{\beta}$), if the burst is
less than 10 Myr old. This spectral index is rather independent of the  metallicity and
IMF. Therefore, any deviation from the predicted spectral index can be
attributed to reddening.

First, the spectrum is corrected for the Galactic extinction, which we derive assuming a
ratio of N$_{HI}$/E(B-V)= $4.93\times10^{21}$ cm$^{2}$ (Bohlin 1975) and an HI column
density towards NGC 7714 of $4\times10^{20}$ cm$^{-2}$ (Stark et al 1992). The color
excess derived is $E(B-V)=0.08$; this implies an extinction of 0.65 mag at 1500 \AA. The
slope measured in the corrected spectrum (log F$_\lambda$ vs. log $\lambda$) after
applying the correction for foreground extinction is $\beta$=-2.45, very close to the
theoretical value of -2.5 which corresponds to the age deduced from the fit to the
absorption profiles. This indicates very low additional internal extinction associated
with the starburst. Using the empirical extinction law from Calzetti, Kinney, \&
Storchi-Bergmann (1994) to match the slope $\beta$=-2.45 to the expected slope
$\beta$=-2.5 at 5 Myr, we derive an internal extinction in the starburst of $E(B-V)=0.03$
(0.26 mag at 1500 \AA). The spectrum corrected for total extinction matches the slope of
the synthetic spectrum that fits the wind absorption lines (see Figure 5). We note that
the total extinction derived from the ultraviolet continuum is lower than the total
extinction derived from the Balmer decrement, E(B-V)=0.21 (GD95). This discrepancy between
the extinction values derived from the two methods has previously been found by  Fanelli,
O'Connell, \& Thuan (1988) and Calzetti, Kinney, \& Storchi-Bergmann (1994).  This result
suggests that probably ionized gas and stars are not co-spatial in a starburst   (Calzetti
1996).

The extinction derived above pertains only to the ultraviolet light within the GHRS
aperture; however, the IUE spatial profile indicates that the ultraviolet light is
extended, at least along P.A.=135$^{\circ}$.45. Figure 6 shows the IUE ultraviolet light
profile that was obtained by adding all the data in the dispersion direction in the
wavelength range from 1250 to 1850 \AA. The IUE aperture was oriented at P.A=
135$^{\circ}$.45. The profile shows a wing on one side that extends between  4 and 10
arcsec from the camera pixel with the highest emission. This wing could represent the
ultraviolet emission from the circumnuclear region called A by GD95. We have followed the
same procedure to determine the extinction for the IUE spectrum. After correcting for
Galactic extinction, the spectral index of the corrected IUE spectrum is $\beta$=-1.1. 
This result is in agreement with that found by Heckman et al (1998), $\beta$=-1.03. The
change of the slope with respect to the GHRS spectrum indicates a change of the extinction
and/or of the stellar population with the aperture. The change of the extinction with the
aperture could be due to the swept dust by the stellar winds and SN explosion. Using the
empirical extinction law from Calzetti et al (1994), we derive an internal extinction
$E(B-V)=0.3$ (2.8 mag at 1500 \AA). This represents an upper limit to the reddening,
because an extended population of B and A stars could contribute significantly to the
ultraviolet continuum light in the IUE aperture, flattening  the spectral slope, in which
case the intrinsic reddening of the starburst would be lower. Thus, the IUE  spectrum
reveals the presence of extended star formation which could be more dusty than the nuclear
burst and/or more evolved.  Alternatively, the IUE aperture could pick up a large flux
contribution from an older  field population, that makes the slope of the spectrum flatter
than expected by a hot  stellar population. Note that the color excess derived depends on
the extinction law used.  If we use the LMC or the SMC law, the E(B-V) derived is 0.2 and
0.1, respectively, instead of 0.3.

\subsection{Intrinsic Luminosities and the WR/O ratio} 

After correcting for the total extinction, E(B-V)= 0.11, the monochromatic luminosity at
1500 \AA\ in the GHRS aperture is $10^{39.9}$ erg s$^{-1}$ \AA$^{-1}$. Assuming that the
ultraviolet flux is due to a burst of 5 Myr, and a Salpeter IMF with upper limit mass
cut-off of 100 M$\odot$, the corresponding number of O stars is about 16600.  Note that an
upper cut-off mass of 40 M$\odot$ gives the same result. These stars produce a photon
ionizing luminosity of $7.9\times10^{52}$ s$^{-1}$. We have measured the nuclear H$\beta$
flux from the 2D data in GD95 within two different apertures of sizes that bracket the GHRS
aperture size, $1.2\times3.3$ arcsec$^2$ and $1.2\times2$ arcsec$^2$. After correcting for
the extinction derived from the Balmer decrement, the ionizing photon luminosity deduced
from the H$\beta$ flux within these two apertures is $8.2\times10^{52}$ s$^{-1}$ and
$6.0\times10^{52}$ s$^{-1}$, respectively. This is in very good agreement with the
ionizing photon luminosity deduced from the UV continuum.  However, some H$\beta$ flux
could be lost due to seeing and the width of the slit. Therefore a conservative upper
limit to the ionizing photon luminosity is 1.4-1.6 times the values derived above. The
estimated mass of the starburst is at least $5\times10^6$ M$\odot$ (integrating the IMF
down to a lower mass limit of  1 M$\odot$), and the bolometric luminosity is $5\times10^9$
L$\odot$, similar to the nuclear starbursts observed in Seyfert 2 galaxies (Gonz\'alez
Delgado et al 1998b).  The ionizing photon luminosity predicted by continuous star
formation models is 10$^{53.5}$ ph s$^{-1}$, a factor 4 larger than the value deduced from
the H$\beta$ flux.
     
We can estimate the number of WR stars from the luminosity of the HeII $\lambda$1640 line
measured within the GHRS aperture. After dereddening,  and assuming the calibration given
by Conti (1996; 1 WN corresponds to $6.3\times10^{36}$ erg s$^{-1}$), the line flux gives
a luminosity of $10^{40.1}$ erg s$^{-1}$, corresponding to about 2000 WR stars. This
number of WR stars gives a WR/O ratio of 0.12. The equivalent width of the HeII
$\lambda$1640 is 2.7 \AA. These two numbers are compatible with the values predicted by an
instantaneous burst (see Figures 2 and 10 in Schaerer \& Vacca 1998). Our conclusion is
that a burst 5 Myr old with a mass of 5$\times$10$^6$ M$\odot$ is able to explain the
continuum and stellar UV absorption features  in the central 1.74$\times$1.74 arcsec of
NGC 7714, with a Salpeter IMF and upper mass limit higher than 40 M$\odot$. 

The monochromatic luminosity at 1500 \AA\ in the IUE aperture is $10^{41.25}$ erg s$^{-1}$
\AA$^{-1}$, after correcting for a total reddening E(B-V)=0.38. This implies a bolometric
luminosity of $1.1\times10^{11}$ L$\odot$,  an ionizing photon luminosity  of
 2$\times$10$^{54}$ ph s$^{-1}$ and a mass of $1.2\times10^8$ M$\odot$ (assuming an age of
5 Myr), a factor 20 larger than the values derived from the GHRS aperture. The IUE
bolometric luminosity is similar to the total IR luminosity, $5\times10^{10}$ L$\odot$,
measured by IRAS and computed with the four wavelength bands as defined by Sanders \&
Mirabel (1997). Thus, this extended star-formation region could contribute significantly
to the overall galaxy  bolometric luminosity.  However, these results have been derived
under the assumption that the  change in the slope in the IUE spectrum is due to internal
extinction. If an underlying older population contributes significantly to the IUE
aperture by flattening the spectrum, the derived color excess, and the mass, bolometric
luminosity and the ionizing photon luminosity, represent an upper limit to the true
values. The other extreme case is to assume that no internal extinction affects the IUE
flux. In this case,  the monochromatic luminosity at 1500 \AA\ is 10$^{40.1}$ erg s$^{-1}$
\AA$^{-1}$. This  implies an ionizing photon luminosity of 1.2$\times$10$^{53}$ ph
s$^{-1}$, which is a factor  2.5 lower than the rate of ionizing photons derived from the
H$\alpha$ image adding the  contribution of the nucleus and the region A (note that these
two components fall within the IUE  aperture,  see Figure 1 in GD95). Thus, the change of slope
is probably due to both, a  change of reddening and the contribution of an underlying
population.

Evidence that an underlying population contributes to the IUE aperture comes from the  CIV
profile. This line is significantly more diluted in the IUE than in the GHRS  spectrum
(Figure 7). Thus, the IUE spectrum suggests that an underlying population contributes  to
its flux, and this makes the determination of the starburst properties  from the IUE flux
rather uncertain.

\subsection{Interstellar lines and L$\alpha$ emission} 
       
The ultraviolet spectrum shows two systems of interstellar lines, one formed in the Milky
Way halo and the other in NGC 7714 (see Figure 1). The most prominent lines in
NGC 7714 are AlII $\lambda$1670, FeII $\lambda$1608, SiII $\lambda$1527, CII
$\lambda$1335, OI+SiII $\lambda$1302, SiII $\lambda$1260 and SiII $\lambda$1190.
The equivalent widths of these lines are shown in Table 1. Their values are uniformly distributed between 2 and 3 \AA, indicating that the lines are saturated. In the saturated part of the
curve of growth the line equivalent width depends on the velocity dispersion
rather than on the column density, so it can be used to constrain the
kinematics of the interstellar medium (Heckman \& Leitherer 1997; Gonz\'alez
Delgado et al 1998a). The equivalent width of the SiII $\lambda$1260 implies a
velocity dispersion of 160 km s$^{-1}$, which is consistent with the upper limit
derived from the FWHM of these lines. Note that the lines are not resolved since
our instrumental resolution is 3.3 \AA, corresponding to a velocity dispersion
of about 300 km s$^{-1}$. The observed equivalent width could be the result of large
scale motions of the interstellar gas. This has also been observed in other
starburst galaxies (Heckman \& Leitherer 1997; Gonz\'alez Delgado et al 1998a).
However, in contrast to other  starbursts, the ultraviolet absorption lines in NGC 7714
are not blueshifted with respect to their systemic velocity. 

L$\alpha$  is observed in emission. The measured flux is
2.9 $\times 10^{-14}$ erg s$^{-1}$ cm$^{-2}$ and the equivalent width is 1.5 \AA. The line
shows a pronounced drop in the blue wing, due to interstellar L$\alpha$ absorption in our Galaxy
and in NGC 7714 itself. This absorption shifts the peak of the emission
line to the red with respect to the systemic velocity. This has also been observed in
other starburst galaxies (Lequeux et al 1995; Gonz\'alez Delgado et al 1998a; Kunth et al 1998). To
correct the L$\alpha$ emission from the absorption, we have performed a multiple
component fit to the observed absorption features (SiII $\lambda$1190,1193, Galactic L$\alpha$ and L$\alpha$ in NGC 7714; we assume that L$\alpha$ in NGC
7714 is at the rest wavelength) using theoretical Voigt profiles produced with
the XVoigt software package (Mar \& Bailey 1995). After dividing the observed
spectrum by the fitted Voigt profiles, the L$\alpha$ flux is $2.9\times10^{-13}$
erg s$^{-1}$ cm$^{-2}$ and the L$\alpha$/H$\beta$ ratio 2.9 (4.1 using the
H$\beta$ flux measured in the central 1.2$\times$2 arcsec$^2$). Correcting by
the reddening derived from the Balmer decrement and using the LMC extinction
law, the ratio is 23 (33); close to the value predicted from the recombination
theory (33, Ferland \& Osterbrock 1985).  We can conclude that attenuation
by dust due to multiple resonant scattering by hydrogen atoms does not  play an important
role in NGC 7714 because the L$\alpha$/H$\beta$ ratio corrected for extinction is close to the recombination value. 

\section{Stellar content in the nuclear starburst: fitting the  optical-near
infrared emission line spectrum} 

The emission line spectrum of a starburst depends on the radiation field from 
the ionizing stellar cluster, and the density and chemical composition of the
gas. In this section, we predict the emission line spectrum using a
photoionization code that takes as input the spectral energy distribution
generated by a stellar evolutionary synthesis code. This technique is applied as
a second independent method to constrain the star formation history and the age
of the nuclear starburst in NGC 7714. It has been previously used to study the
stellar content in starbursts and giant HII regions (Garc\'\i a-Vargas \& D\'\i
az 1994; Garc\'\i a-Vargas, Bressan, \& D\'\i az 1995a, 1995b; Stasi\'nska  \&
Leitherer 1996; Garc\'\i a-Vargas et al 1997). The goal is to perform a test of
consistency between the young stellar content derived from the ultraviolet absorption
spectrum and from the optical emission lines. First, we describe the basis of
the  stellar evolutionary synthesis and photoionization models.

\subsection{Stellar evolutionary synthesis models} 

The spectral energy distribution was generated by the evolutionary synthesis
code  developed by Leitherer and collaborators (Leitherer, Robert, \& Drissen
1992). The code has been updated recently. A description of the new version
will be presented elsewhere (Leitherer et al 1998, in preparation). Two
important changes have been made with respect to the previous version of the
code. The most recent stellar evolutionary models of the Geneva group
(Schaller et al 1992; Maeder 1994), and the stellar atmospheres grid compiled by
Lejeune et al (1996) are used, instead of Maeder's (1990) and Kurucz's models
(1992). As in the previous version of the code, we use the expanding spherical
extended non-LTE models published by Schmutz et al (1992) for stars with very
strong winds. 

The spectral energy distribution was generated using the 0.4 Z$\odot$ metallicity
tracks (note that the chemical abundance derived from the emission lines in NGC
7714 is close to half solar, GD95), and assuming two different star formation
scenarios (instantaneous burst and continuous star formation). In both cases, a
Salpeter IMF with an upper and lower mass limit cut-off of 80 M$\odot$ and 1
M$\odot$, respectively, was assumed. 

\subsection{Photoionization models }

We take the spectral energy distribution as input to the photoionization code
CLOUDY (version 90.04, Ferland 1997). CLOUDY resolves the
ionization-recombination and heating-cooling balances, and predicts the
ionization structure of the nebula, the temperature, and the intensities of the
emission lines. We assume that the nebular gas is ionization bounded and
spherically distributed around the ionizing cluster with a constant density.  We
assume an inner radius of 3 pc, but the outer radius is determined by the
ionization front. The chemical composition of the gas is scaled to the oxygen
abundance (close to half solar), except for He, S, Ne and N. The abundances and
the electron density were determined from the optical emission lines as given by GD95. 
They are shown in Table 2. The adopted values of the solar
abundance are as in Stasi\'nska (1990), except the sulfur
abundance, for which we take the S/O ratio given by Grevesse \& Anders (1989).
The ionizing photon luminosity is fixed to log(Q)=52.9 ph s$^{-1}$,
as derived above independently from the Balmer recombination lines and from
the ultraviolet continuum luminosity. Models are computed taking the
filling factor as a free parameter, 
with values of $10^{-1}$, $10^{-2}$, $10^{-3}$, $10^{-4}$, and $10^{-5}$.
The change
in the filling factor is equivalent to changing the ionization parameter U,
defined as Q/(4$\pi$R$_s$N$_e$c); where Q is the ionizing photon luminosity,
N$_e$ the electron density, c the speed of light and R$_s$ the Str\"omgren
radius. The average U is proportional to ($\phi^2$ N$_e$ Q)$^{1/3}$, where $\phi$
is the filling factor. 

The emission line spectrum depends on the ionizing radiation field --- 
i.e. on the evolution of the cluster --- and on the ionization 
parameter for a fixed geometry,
 metallicity, density and the total number of ionizing
photons. First, we
determine the ionization parameter using the ratio
[SII]$\lambda$6716+6731/H$\beta$. This ratio is a good calibrator of U for
continuous star formation and for burst models (Figure 8). The observed ratio
[SII]$\lambda$6716+6731/H$\beta$=0.58 indicates an ionization parameter of -2.9,
corresponding to a filling factor of 0.001. Then, we use the other emission lines
to constrain the star formation scenario (burst or continuous star formation) and
the age of the starburst. Figure 9 shows the line ratios
[OIII]$\lambda$5007/H$\beta$,  [SIII]$\lambda$9069/H$\beta$,
[OII]$\lambda$3727/H$\beta$, [NII]$\lambda$6584/H$\beta$,
[OI]$\lambda$6300/H$\beta$ and HeI$\lambda$4471/H$\beta$ as a function of
age for continous star formation and burst models with filling factor 0.001.  Burst models predict
line ratios that are in better agreement with the observed ratios than continuous
star formation models.  Table 3 shows the strength of the emission lines observed and
the values predicted by the 4.5 Myr burst. The 4.5 Myr burst model fits well the strongest
emission lines, but it does not reproduce the [OIII]$\lambda$4363 and [OI]$\lambda$6300 lines;
the values observed are a factor of two larger than predicted. GD95 did
 three different estimations of the electron temperature of the nebula using the ratio of  
the auroral to the nebular collisionally excited lines [SIII], [SII] and [NII] lines. The 
three ratios give a similar electron temperature (about 7500 K), and with this electron
temperature the metallicity was derived. However, the observed 
[OIII]$\lambda$4363/ [OIII]$\lambda$5007 ratio is larger than the expected ratio for a 
nebula with a electron temperature of 7500 K. This is probably a consequence of a larger 
contribution of the  [OIII]$\lambda$4363 line to the temperature ratio. In a nebula the 
presence of shocks can significantly  affect the lines  [OIII]$\lambda$4363 and  
[OI]$\lambda$6300 (Peimbert, Sarmiento, \& Fierro 1991). Shocks are naturally expected to 
be present in the ionized gas surrounding a starburst with age of 5 Myr as a result of 
the stellar wind and the explosion of the first supernovae in the stellar cluster. At this 
age, the deposition rate 
of mechanical energy is about 10$\%$ of the ionizing luminosity. Thus, shock heating 
could probably account for the excess emission in [OIII]$\lambda$4363 and  [OI]$\lambda$6300
 with respect to the photoionization values.

This photoionization model predicts a Str\"omgren
radius of 185 pc. If the nebula is radiation 
bounded, the radius is a function of the ionizing photon flux, the electron density and 
the filling factor. Then for fixed Q, Ne and $\Phi$, the radius has to be equal to the size of the nebula. We know 
that the ionizing cluster is not a point source because the 
ultraviolet interstellar absorption lines in the GHRS spectrum are not resolved. It is 
at least as extended as the 
size of the GHRS aperture (330$\times$330 pc). This size is similar to the Str\"omgren radius 
predicted. Also, from the WFPC2 image (see Figure 2), we estimate that the size
 of the nebula is about 210 pc. Thus, the predicted size of the Str\"omgren
radius is compatible with the estimated size of the HII nebula. This result is
also consistent with the HII region being radiation bounded. Further evidence comes from 
the [OI]$\lambda$6300 line. This low ionization line forms in the partially 
ionized zone, which does not exist in a matter bounded nebula. On the other hand,
 the ionizing photon flux derived from the ultraviolet continuum luminosity is very 
similar to that derived from the H$\beta$ luminosity, indicating again that the radiation
 bounded hypothesis is correct. 

Thus, the conclusion is that emission line ratios can discriminate very well between the burst and 
continuous star formation (csf) scenario. The csf models can be excluded because they predict emission line ratios which are higher than the values observed (except for the [NII] $\lambda$6584/H$\beta$). The best fit is found for a burst 4.5 Myr old. Is it reasonable that all these stars are formed in an instantaneous burst?
The ultraviolet data indicate that the starburst is extended at least 1.74 arcsec (330 pc). 
The HST optical image shows that it may not be more extended than 400 pc (in diameter). Taking 400 pc as
 the diameter of the nuclear starburst and the velocity dispersion for this galaxy of 
180 km s$^{-1}$ (GD95), we derive that the crossing time within the starburst is 2.1 Myr. 
This time is significantly shorter than the evolutionary lifetime of a single generation of
 massive stars. Thus, it is dynamically possible that the 
nuclear starburst in NGC 7714 formed in an instantaneous burst of star-formation because the crossing-time is shorter than the duration of the starburst.

\section{Discussion}

\subsection{X-ray and radio emission}

Can the derived stellar population explain the observed X-ray luminosity and 
radio continuum emission? X-ray emission in normal galaxies can have several
origins. As discussed by Martin \& Kennicutt (1995) and Stevens \& Strickland
(1998), production of X-rays in starburst galaxies by AGNs, static galactic
coronae, single stars, or X-ray binaries is rather unlikely to be the 
dominant mechanism. The most plausible interpretation is due to powering by
supernovae, with some contribution from stellar winds.

The models of Leitherer \& Heckman (1995) predict
a type II supernova rate of about 0.007~yr$^{-1}$ for a bolometric luminosity
of $5 \times 10^{9}$ L$\odot$. If we assume that the kinetic energy released per
supernova is $10^{51}$~erg, the mechanical luminosity released by stellar
winds and supernovae is $2 \times 10^{41}$~erg s$^{-1}$. If thermalization
occurs, this luminosity is {\em in principle} available to create a hot,
over-pressured bubble. In practice, the sum of the thermal and kinetic energy
of the bubble will be lower due to radiation losses. Hydrodynamical simulations
by Thornton et al. (1998) suggest radiation losses on the order of 90\%. 
The observed X-ray luminosity in the center of NGC~7714 is 
$6 \times 10^{40}$~erg~s$^{-1}$ (W81). If interpreted as due to thermal emission
from hot, shocked material, it could be accounted for by supernovae
and winds, even allowing for significant radiation losses. 

The observed 20~cm
radio flux is 20~mJy (W81). In the standard supernova driven outflow model
(e.g., Chevalier 1982; 1991) radio emission is assumed to be synchrotron 
radiation from electrons accelerated by shocks in the interaction zone.
Models for the non-thermal component of a hot bubble are substantially more
uncertain than the thermal (X-ray) prediction. Empirical data for individual
local supernova remnants indicate non-thermal luminosities of about 1\% the
thermal value (Chevalier 1982). Assuming the 20~cm flux 
is purely non-thermal, and adopting a bandwidth of $\Delta \nu = 10$~GHz, the
non-thermal component in NGC~7714 has a luminosity of 
$3.7 \times 10^{38}$~erg~s$^{-1}$. Remarkably, this is about 1\% of the 
X-ray luminosity. Therefore we conclude that both the thermal and the 
non-thermal component of the hot, central ISM in NGC~7714 are consistent with
being powered by winds and supernovae.
  
While simple energy considerations are instructive, more detailed 
hydrodynamical calculations for the behavior of a hot superbubble power by
a stellar population are needed. Although still deficient in some aspects
(e.g. with respect to the time dependence of the energy input), the relations
discussed by Martin \& Kennicutt (1995) and Stevens \& Strickland (1998) 
provide further insight. These authors used the standard superbubble model
to solve the energy and momentum equations for an adiabatic bubble.
According to these models, the X-ray luminosity
can be expressed in terms of the starburst luminosity and age for a given
IMF.  A 5~Myr old starburst with a standard Salpeter IMF is predicted to
have a ratio of the X-ray over bolometric luminosity of $\sim$ 10$^{-3}$
(see Fig.~16 of Stevens \& Strickland). Applied to NGC~7714, these models
predict an X-ray luminosity of $2 \times 10^{40}$~erg~s$^{-1}$, within a 
factor of 3 of the observed value. The fact that the prediction is lower
than the observation might suggest a clumpy ISM. An ambient medium containing 
clouds or filaments can generate
extra mass, and therefore extra X-ray luminosity for fixed mechanical
luminosity input (Martin \& Kennicutt). However, the theoretical and 
observational uncertainties are large enough that the data are consistent  
with a homogeneous medium as well.
 
The stellar population, as derived independently from nebular emission lines
and stellar absorption lines, is capable of generating the mechanical energy
input to account for the observed thermal X-ray and non-thermal radio emission
as well. The associated supernova rate is 0.007~yr$^{-1}$. This rate is
consistent with the upper limit imposed by the null result in the supernova
search of Richmond, Filippenko, \& Galisky (1998). No supernovae were detected
in NGC~7714 during a several year long monitoring period. The non-detection
implies an upper limit to the supernova rate of $< 0.1$~yr$^{-1}$. Few if any
supernovae should escape detection due to dust extinction since the extinction
is known to be low. We can account for most of the luminosity at 1500~\AA, which
traces precisely the progenitors of type II supernovae ($M \approx 20$ M$\odot$).

The derived supernova rate is two orders of magnitude lower than that found 
by W81. Their estimate was based on using X-ray and radio luminosities and 
lifetimes
of local supernova remnants as templates and scaling them to the
luminosities observed in NGC~7714. In retrospect this approach has turned out
to be too simple. Supernovae and supernova remnants in dense media, like in
starbursts, can have rather different properties. This may be particularly
relevant to the very young starburst in NGC~7714. Few red supergiants have
formed at an age of 5~Myr, and most of the supernova progenitors are 
predicted to be WR stars. WR stars have hydrogen-free
envelopes and have stellar-wind properties quite different from red supergiants.
On the other hand, W81 use the truncated starburst models of Rieke et al (1980) 
to estimate the mass and rate of supernovae. In these models the IMF is truncated
 with the high mass $\leq$ 30 M$\odot$. Therefore, they have to boost the star 
formation rate in oder to get the ionizing photons, and therefore the supernova rate 
became high. However, our data are compatible with a normal IMF.  

In the last few years, young compact supernovae have been studied in detail
at radio wavelengths (Weiler et al, 1986, 1990, 1991, 1992; 
Van Dyk et al 1992, 1993a,b; Ryder et al 1993; Yin 1994). These
supernova, at the peak of their emission, can have a radio luminosity which 
is 100~--~1000 times larger than Cas~A. The presence of these supernovae 
and their remnants can explain the radio luminosity detected in highly 
luminous IRAS galaxies (Colina \& P\'erez-Olea, 1992; 
P\'erez-Olea \& Colina, 1995).  Evidence that these compact 
supernova can account for the radio emission comes from 
18~cm VLBI continuum imaging observations of Arp~220. The images have 
revealed a dozen compact radio sources, with fluxes 
that can be explained by compact young radio supernova (Smith et al 1998). 

Assuming that the 6 cm radio emission observed by W81 is non-thermal emission emitted by
the central 1.74$\times$1.74 arcsec, and using the models of Colina \& P\'erez-Olea
(1995), we derive a supernova rate of 0.07 yr$^{-1}$, which is in agreement with our
empirical upper limit determination, but still one order of magnitude higher than the rate
predicted by the model which accounts for the UV luminosity.  To predict a supernova rate
as high as 0.07 yr$^{-1}$, we need a central starburst with a UV luminosity higher than
10$^{39.9}$ erg s$^{-1}$ \AA$^{-1}$. Is the estimated luminosity wrong because of a wrong
estimation of the intrinsic reddening derived from the UV continuum slope? If the internal
reddening were $E(B-V)=0.25$, then the UV luminosity of the starburst in the GHRS aperture
would be 10$^{40.7}$ erg s$^{-1}$ \AA$^{-1}$, the supernova rate 0.05 yr$^{-1}$, the
thermal emission 4 mJy, and the non-thermal emission 6 mJy. However, by correcting the
GHRS spectrum by $E(B-V)=0.25$, the UV spectrum appears clearly over-corrected when
compared with the synthetic model. From exploring the fit of the UV continuum slope, we
estimate that the intrinsic reddening cannot be higher than 0.1. Correcting for this
extinction, the luminosity at 1500 \AA\ is 10$^{40.2}$ erg s$^{-1}$ \AA$^{-1}$, the mass
of the starburst  1.1$\times$10$^7$ M$\odot$ and the supernova rate 0.02 yr$^{-1}$. In
this case, we can account for just one third of the radio emission.  Compact supernova
remnants are also very strong X-ray sources with luminosities in the range
10$^{38}$-10$^{40}$ erg s$^{-1}$ (e.g. SN 1978K, Ryder et al 1992; SN 1986J, Bregman \&
Pildis 1992; SN 1993J, Schlegel 1994). SN 1988Z was detected  from ROSAT HRI observations
with a luminosity of 10$^{41}$ erg s$^{-1}$ (Fabian \& Terlevich 1996). Thus, these
compact supernova remnants are brighter by several orders of magnitude than the Galactic
remnants (10$^{35}$-10$^{37}$ erg s$^{-1}$ range), and they can contribute significantly
to the X-ray flux.

Probably even more important, a starburst nucleus is not just a scaled-up
version of an individual supernova remnant. The superposition of numerous
supernova explosions and the effects of stellar winds create a hot,
over-pressured bubble which expands and can experience a break-out from the
galaxy (Heckman 1995). The relation between the X-ray luminosity of the bubble
and the properties of the stellar population, including the supernova rate, 
can be quite different from what is expected on the basis of individual
supernova remnants (although the physics is related). The success of the
simple superbubble model discussed above is encouraging. We conclude that
the previously derived supernova rate of $\sim$1~yr$^{-1}$ is not supported
by more recent starburst models and that the current supernova rate in the
nucleus of NGC~7714 is close to about 1 per century. 

Incidentally, this rate is about the same as the disk-integrated rate observed
in normal spirals, including the Milky Way. NGC~7714 does not have a 
significantly higher total supernova rate than observed in spirals with
active star formation. What distinguishes NGC~7714 (and other nuclear 
starbursts), is the concentration of the supernova events in the nucleus.
The {\em nuclear} supernova rate per unit surface is about three orders of 
magnitude higher than that of normal galaxies. 

\subsection{Evidence of an additional stellar population}

Is there evidence for an additional stellar component younger than 30-40 Myr in the nucleus of NGC 7714?  
Bursts younger than 30-40 Myr will contribute to the non-thermal emission (stars more massive than 8 M$\odot$ explode as supernovae, and their life-time is less than 30-40 Myr).  When 
older than 10 Myr, they will not  contribute to the strength of the emission lines, but will affect the equivalent widths of the recombination emission 
lines. For a burst 4.5 Myr old, the equivalent width of H$\beta$ is 65 \AA, however, the observed 
value is 30 \AA. Thus, the equivalent width of H$\beta$ must be diluted. 
But we do not know whether it is diluted by a burst younger than 30-40 Myr, 
or by the bulge population. In the near-infrared, the spectrum shows the CaII triplet in absorption. The strength of these lines is dependent on both metallicity and gravity 
(D\'\i az, Terlevich, \& Terlevich 1989). At solar metallicity, these lines are very strong in starbursts if the near-infrared light is dominated by red supergiant stars. Population synthesis models of the two strongest absorption lines (CaII $\lambda$8542 and CaII $\lambda$8562) for half solar metallicity predict an equivalent width for the two lines of 4-5 \AA\ for burst younger than 40 Myr (Garc\'\i a-Vargas et al 1998). The observed value is 4.5 \AA\ (GD95), in agreement with the prediction of the models. However, the same models predict an equivalent width somewhat larger than 5 \AA\ for a a burst older than 100 Myr. Thus, even though the strength of the CaII triplet is compatible with a burst younger than 40 Myr, we cannot exclude that the lines could be dominated by red giant instead of red supergiant stars. Thus, the equivalent width of H$\beta$ and the strenght of the CaII triplet show evidence that an underlying population contributes to the central 1.7$\times$1.7 arcsec spectrum of NGC 7714. Further evidence comes from the NIR spectrum. We will present a full analysis of the UV-optical-NIR continuum in paper II (Goldader et al 1998, in preparation).

\section{Summary and conclusions}

The main goal of this work is to constrain the most recent star formation
history in the central 300 pc of NGC 7714. To archieve this goal, we observed the nucleus of NGC 7714 from the
UV to the near infrared. Consistent with previous observations, our data show that NGC 7714 
exhibits the spectral dichotomy typically observed in starburst galaxies. The ultraviolet 
spectrum is dominated by absorption features formed in the intestellar medium and stellar
 winds, while the optical spectrum is dominated by emission lines formed in the photoionized interstellar
 medium. We constrain
the most recent star formation history in the nucleus of NGC 7714 by applying two different
techniques, one based on the ultraviolet spectrum, and the other on the optical to near
infrared emission line spectrum. 

Evolutionary synthesis line profiles of the CIV
and SiIV wind lines are used to constrain the duration of the starburst, the age of the 
stellar population, and the IMF parameters
(slope and upper limit). We find that the profiles of the wind lines are compatible
with a single burst, an age of 5 Myr and a Salpeter IMF with upper mass cut-off higher than 40 M$\odot$. Arguments of causality suggest that is dynamically possible that the nuclear starburst formed an instantaneous burst since the dynamical crossing time within the starburst is lower than the evolutionary life time of massive stars. Evolutionary
synthesis models coupled to the photoionization code CLOUDY are also used to
generate the emission line spectrum. Continuum star formation models are not consistent with the strength of the emission line ratios. The best fit to the emission lines is found
for a 4.5 Myr burst model. From the ultraviolet continuum luminosity we derive a mass of
the starburst of 5--11$\times10^6$ M$\odot$ (note that this probably a lower limit to the mass of the starburst because it has been derived from the UV continuum luminosity) and a bolometric luminosity of
5--10$\times10^{9}$ L$\odot$, assuming stars down to 1 M$\odot$ are formed. Very little extinction is associated with this nuclear
starburst. WR
features at ultraviolet (HeII $\lambda$1640) and optical (HeII $\lambda$4686) wavelengths
are detected. Their luminosities imply a WR/O ratio of  $\sim0.1$, which is
also compatible with the predicted values from the burst models. 

This model predicts a supernova rate of 0.007--0.02 yr$^{-1}$, consistent with the rate derived 
from the radio and the X-ray luminosities, 0.05 yr$^{-1}$ (if the
 emission  accounts for the contribution of radio and X ray compact supernova). This rate is 
in agreement with an upper limit of 0.1 yr$^{-1}$ based on the non-detection of supernova events 
in the past several years. This supernova rate is two order of magnitude lower than previously 
suggested. The supernova rate in the NGC 7714 nucleus is similar to disk-integrated rates in normal
 spiral galaxies.

{\bf Acknowledgments}

We thank Gary Ferland for kindly making his code available, and David Mar for
providing his XVoigt software package. M.L. Garc\'{\i}a-Vargas is grateful to
the Space Telescope Science Institute for the kind support provided during part
of  this work. We have benefited from stimulating and helpful
discussions with Itziar Aretxaga, Miguel Cervi\~no, Luis Colina, Ariane Lan\c{c}on, Enrique P\'erez, Daniel Schaerer and Kim Weaver. We thank the staff at STScI for their help in obtaining the data
presented in this paper and to Antonella Nota for her help at the beginning of this project.
 This work was supported by HST GO-06672.01-95A from the Space Telescope
Science Institute, which is operated by the Association of Universities for
Research in Astronomy, Inc., under NASA contract NAS5-26555.

\begin{deluxetable}{lcc} \footnotesize \tablecaption{Equivalent width (\AA) of
the absorption and emission  features\tablenotemark{a}} \tablewidth{0pt}
\tablehead{ \colhead{Line} & \colhead{type} & \colhead{Ew (\AA)}    }  \startdata

NV $\lambda$1240      & W  & 2.8 \nl NV $\lambda$1240      & W  &-1.2 \nl SiII
$\lambda$1260    & IS & 1.7 \nl OI+SiII $\lambda$1300 & IS & 3.1 \nl CII
$\lambda$1335     & IS & 2.8 \nl SiIV $\lambda$1400    & W+IS & 6.4 \nl SiIV
$\lambda$1400    & W  &-0.2 \nl SiIII $\lambda$1417   & Ph & 0.6 \nl CIII
$\lambda$1428    & Ph & 0.5 \nl SV $\lambda$1501      & Ph & 0.8 \nl SiII
$\lambda$1526    & IS & 2.1 \nl CIV $\lambda$1550     & W+IS & 9.4 \nl CIV
$\lambda$1550     & W  &-0.9 \nl FeII $\lambda$1608    & IS & 1.8 \nl HeII
$\lambda$1640    & W  &-1.6 \nl \enddata  \tablenotetext{a} {IS means
Interstellar, W wind, and Ph Photospheric line. Negative values are lines in
emission.}  \end{deluxetable}

\begin{deluxetable}{lc} \footnotesize \tablecaption{Input parameters for CLOUDY 
\tablenotemark{a}} \tablewidth{0pt} \tablehead{ \colhead{parameter} & \colhead{value}   } 
\startdata

 log N$_e$  &  2.75 \nl
 12+log O/H &  8.53 \nl
 log He/H   &  -1.00 \nl
 log S/H    &  -5.05 \nl
 log Ne/H   &  -4.30  \nl
 log N/H    &  -4.15 \nl

\enddata \tablenotetext{a} {The electron density and metallicity are from GD95, 
and they were derived from the optical emission lines.} \end{deluxetable}

\begin{deluxetable}{lcc} \footnotesize \tablecaption{Line ratio of the emission
lines with respect to H$\beta$   \tablenotemark{a}} \tablewidth{0pt} \tablehead{
\colhead{Line} & \colhead{observed} & \colhead{model}    }  \startdata

 [OII] $\lambda$3727  & 1.97  &  2.13  \nl
 [OIII] $\lambda$4363 & 0.01  &  0.005 \nl
 HeI $\lambda$4471    & 0.037 &  0.04  \nl
 [OIII] $\lambda$5007 & 1.59  &  1.83  \nl
 HeI $\lambda$5876    & 0.12  &  0.12  \nl
 [SIII] $\lambda$6312 & 0.007 &  0.01  \nl
 [OI] $\lambda$6300   & 0.043 &  0.022 \nl
 [NII] $\lambda$6584  & 1.22  &  1.37  \nl
 [SII] $\lambda$6716  & 0.29  &  0.32  \nl
 [SII] $\lambda$6732  & 0.28  &  0.32  \nl
 [SIII] $\lambda$9069 & 0.27  &  0.32  \nl \enddata \tablenotetext{a} {The
observed line ratios are from GD95  corrected for reddening with
C(H$\beta$)=0.3. The line ratios are predicted using the spectral energy
distribution of the 4.5 Myr burst model, assuming a Salpeter IMF and $\rm
M_{up}=80~M\odot$, as the ionizing radiation field into CLOUDY.}
\end{deluxetable}

\clearpage

\figcaption{The 1.74$\times$1.74 arcsec$^2$ GHRS spectrum of
the  nucleus of NGC 7714 (in the rest frame of NGC 7714 wavelength). The most important wind, photospheric and interstellar
absorption lines are labelled. The interstellar absorption lines formed in the
halo of our Galaxy are also labeled as Gal.}

\figcaption{The WFPC2 F606W image. The orientation is North up and
East to the left. The scale is 0.046 arcsec/pix. (a) The image shows
the morphological characteristics of almost the whole galaxy. (b) The central 
2.5$\times$2.5 arcsec$^2$. The nucleus is clearly extended, and resolved into
severals knots.}

\figcaption{$\chi^2$ parameter of the fits to the profile of SiIV for
burst models (a,b), and for the continuous star formation models (c,d) with
different M$_{up}$ (a,c) and IMF slope (b,d). The size of the bubble is
proportional to $\chi^2$, with a larger size indicating a worse fit. The scale
in the small boxes indicates the value of $\chi^2$.}

\figcaption{The observed profiles (dashed line) of CIV (a) and SiIV
(b) and the 5 Myr burst model (thick line), assuming a Salpeter IMF and  $\rm
M_{up}=100~M\odot$.}

\figcaption{The GHRS spectrum dereddened by E(B-V)=0.08 using the MW
extinction law, and by E(B-V)=0.03 using the Calzetti et al (1994) extinction
law, and the synthetic 5 Myr burst model (in relative units). The IMF slope is
Salpeter and $\rm M_{up}=100~M\odot$.} 

\figcaption{The IUE intensity profile perpendicular to the dispersion
direction integrating from 1250 to 1850 \AA. The zero in the X-axis is the IUE
camera pixel with highest ultraviolet emission. The bump between 4 and 9 arcsec could be
the emission associated with the circumnuclear region called A in GD95.} 

\figcaption{The GHRS (thick line) spectrum rebinned to the resolution of the IUE spectrum (thin line). The GHRS flux at 1500 \AA\ is also scaled to match the IUE flux. The spectra are corrected only by Galactic extinction.}

\figcaption{Predicted emission line ratio
[SII]$\lambda$6716+6732/H$\beta$ as a function of the average ionization
parameter generated with CLOUDY, taking as input the spectral energy
distribution from the evolution of a burst (open symbols) and
continuous star forming regimes (full symbols). The observed value is indicated 
by a horizontal line. See text for details.}

\figcaption{Predicted emission line ratios as a function of age. The
filling factor is 0.001. The observed values are indicated by a horizontal line. 
These ratios are compatible with a burst 4.5 Myr old, and they  exclude the continuous star formation 
 models.}

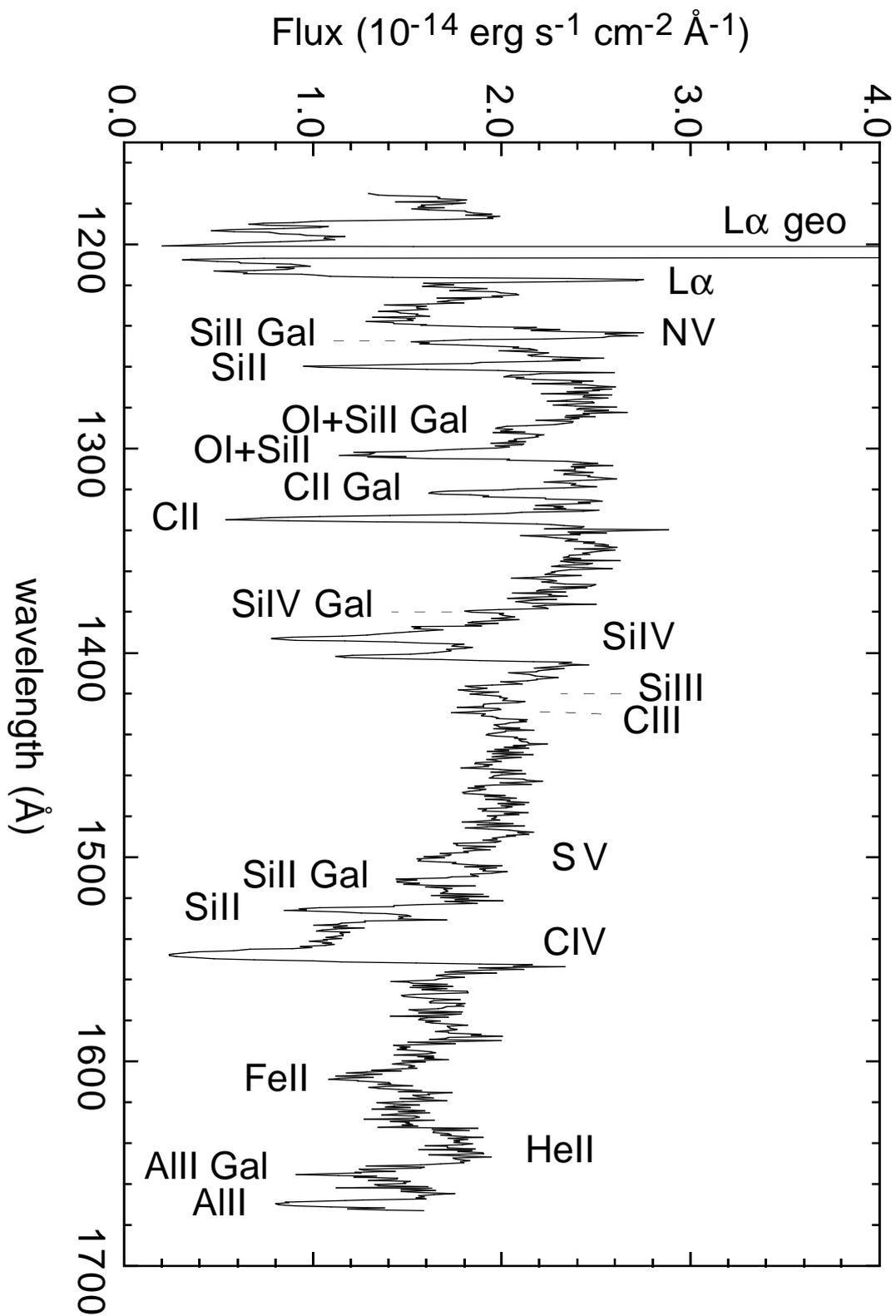
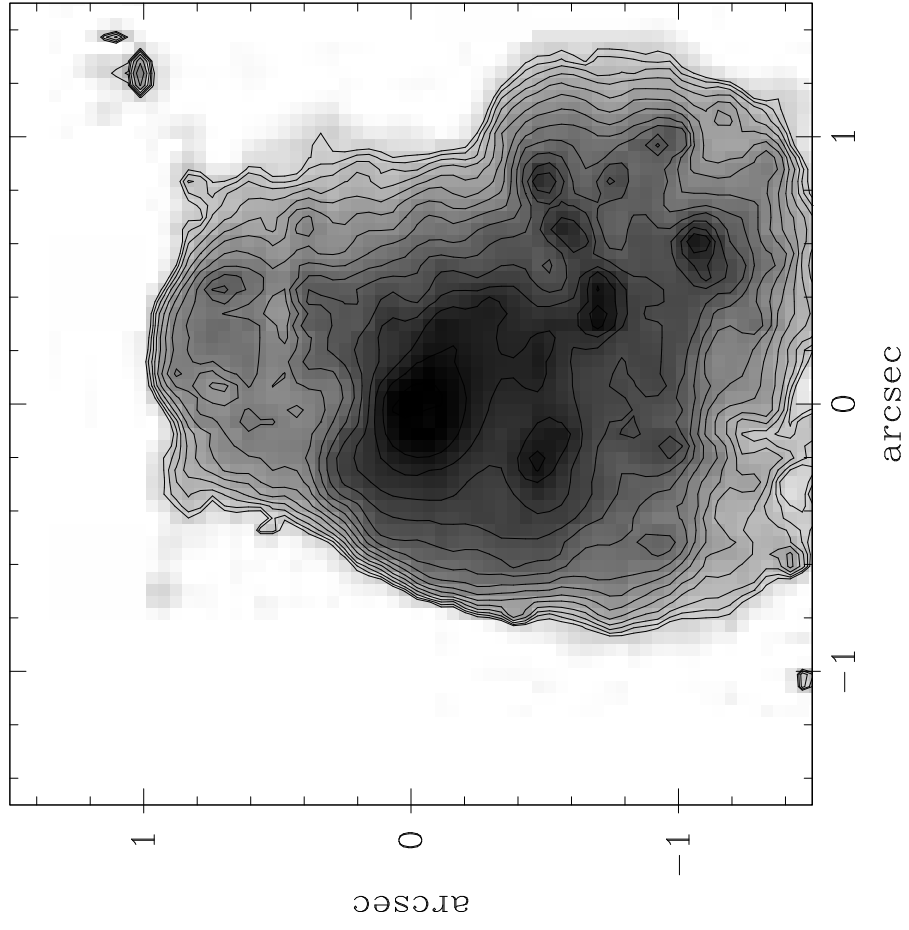
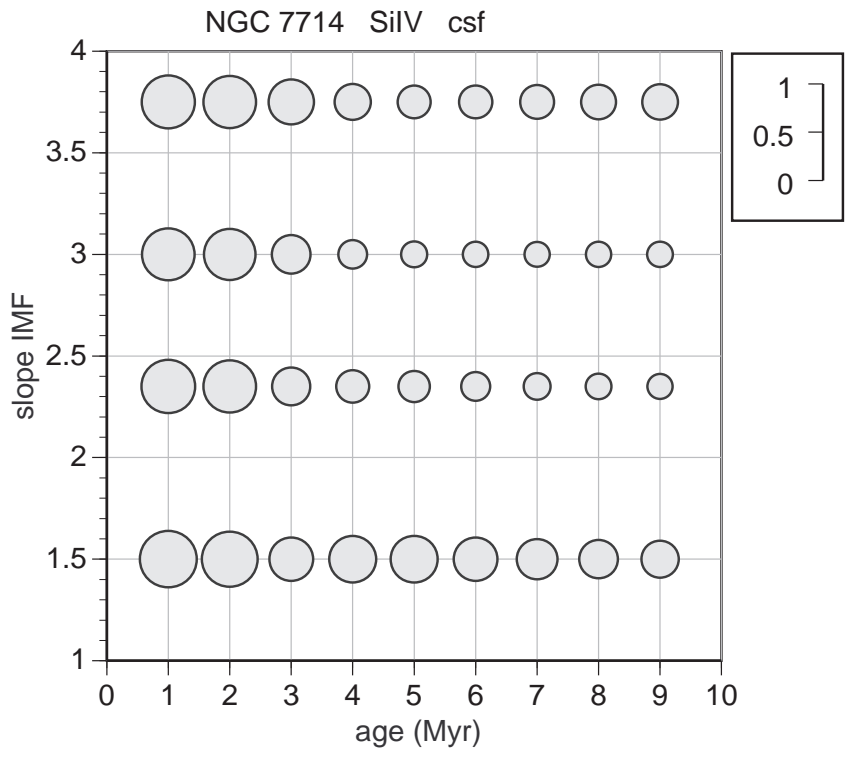
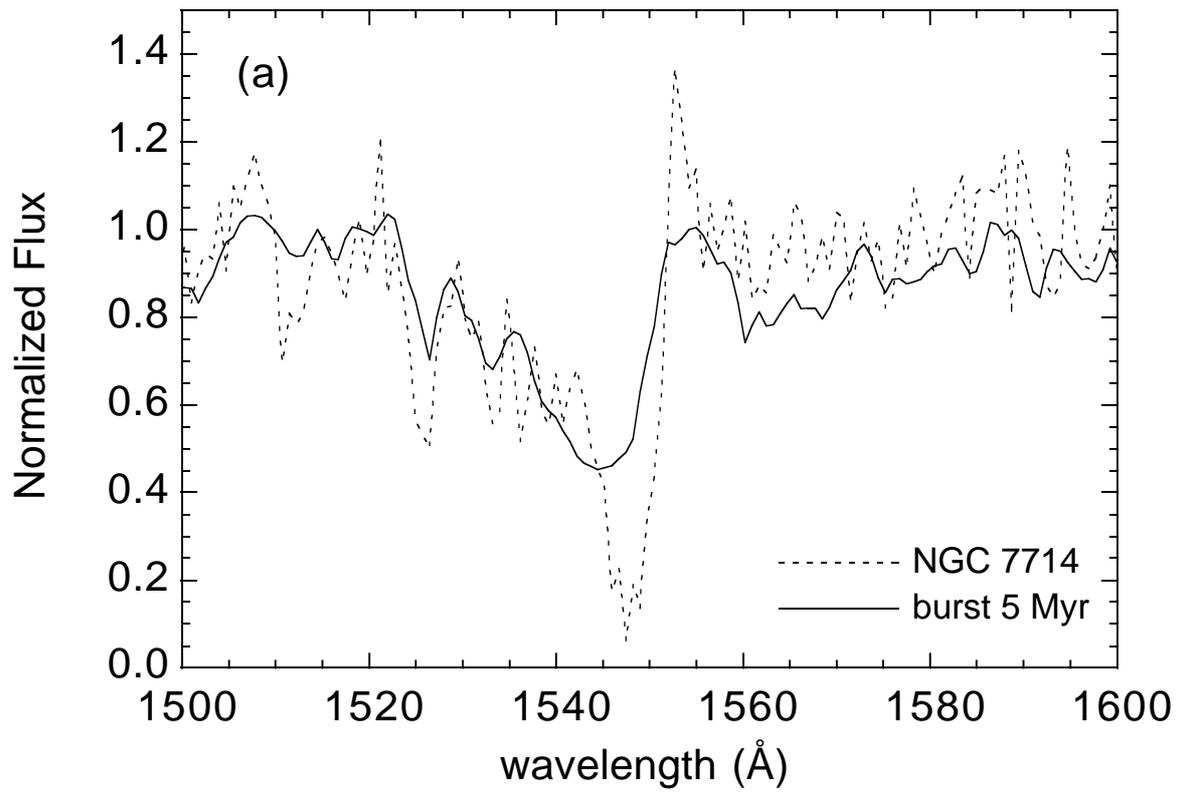
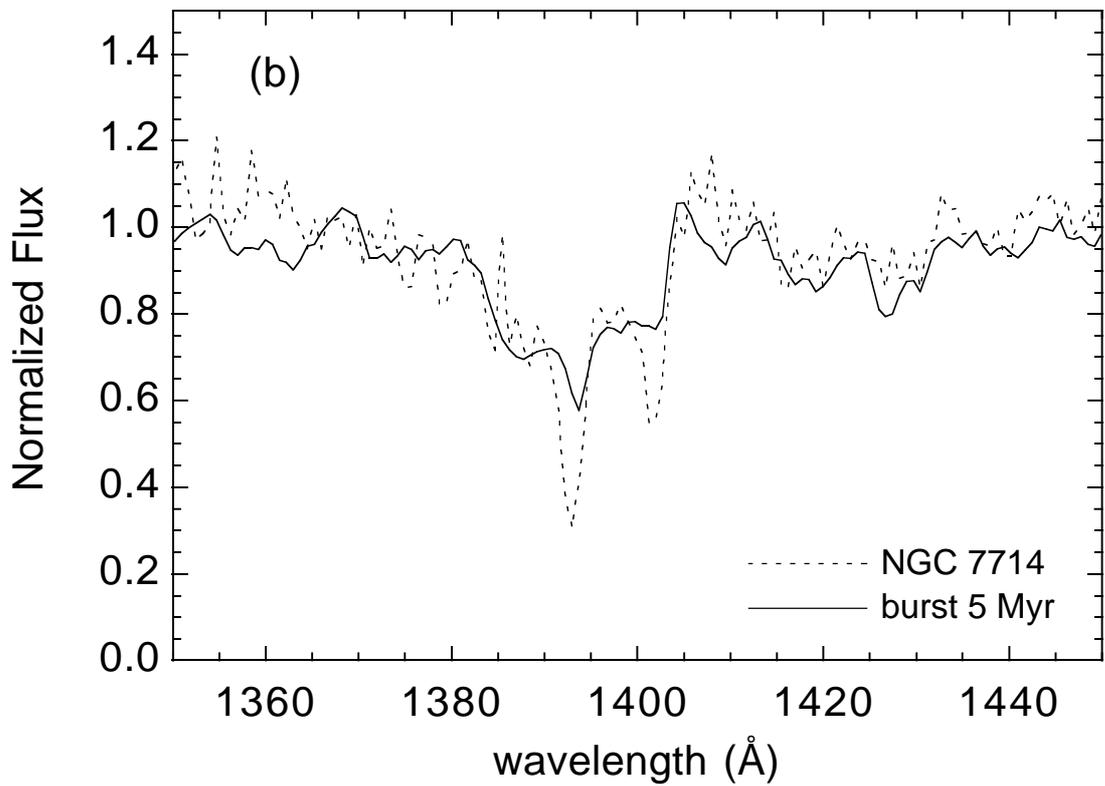
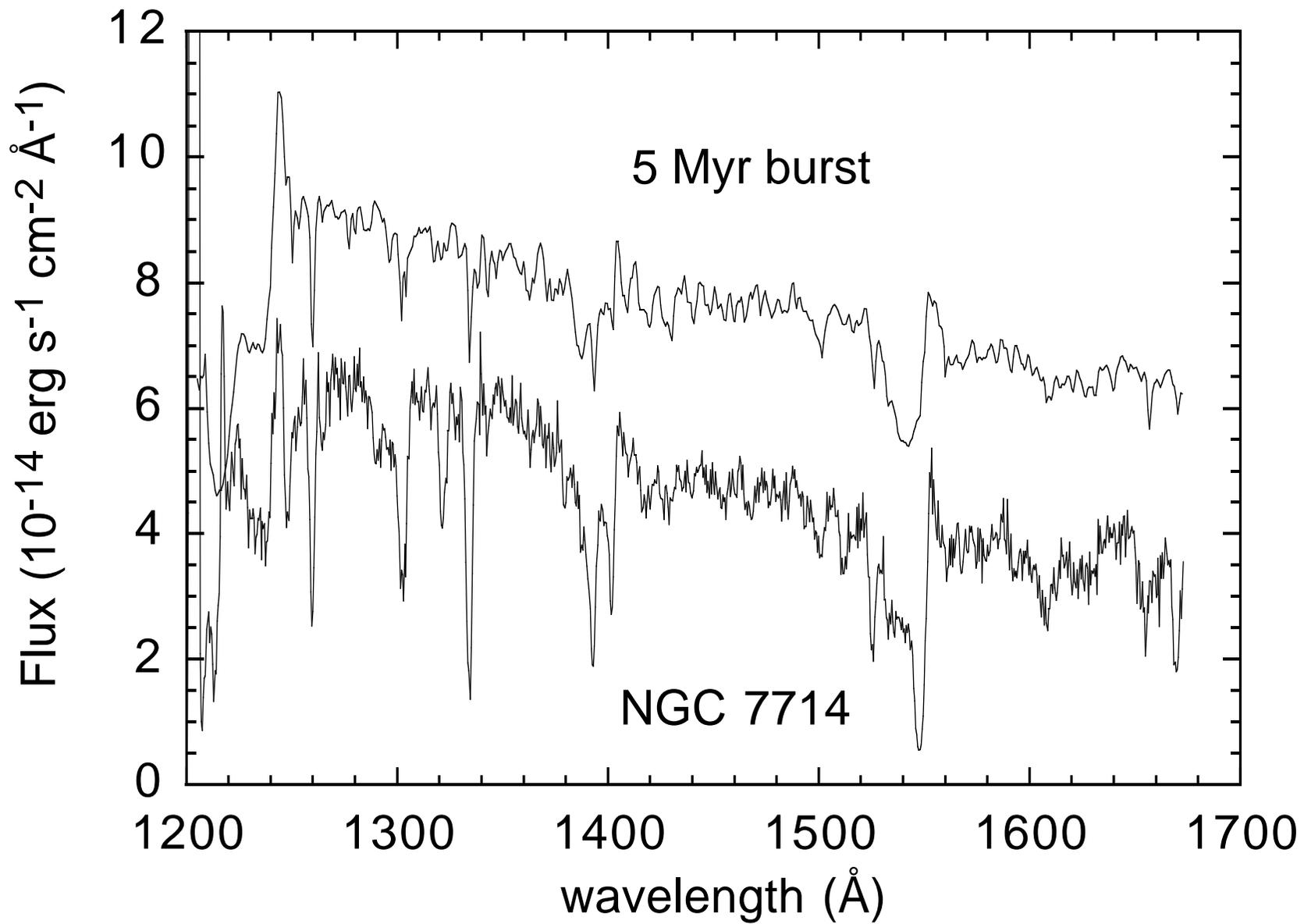
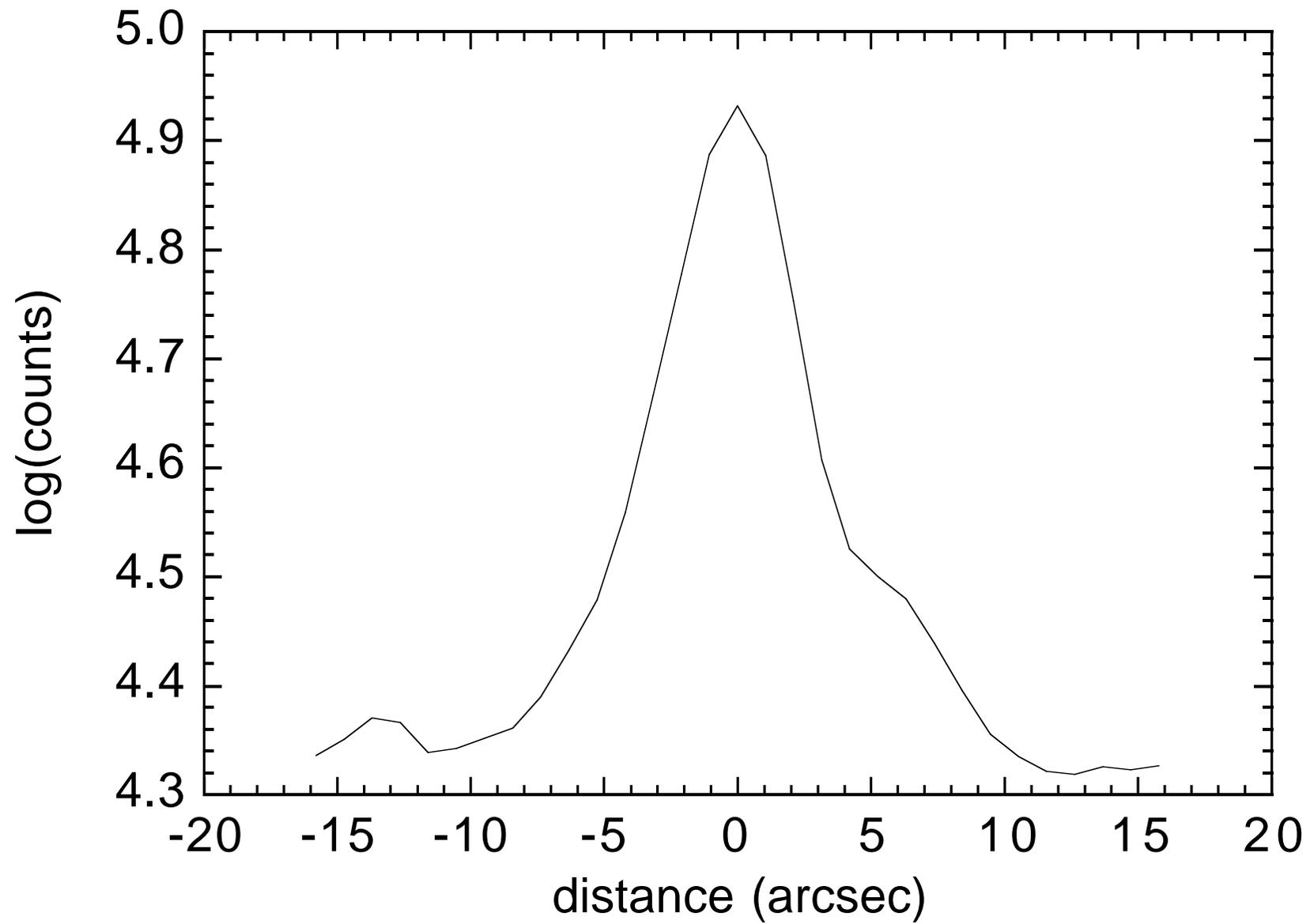
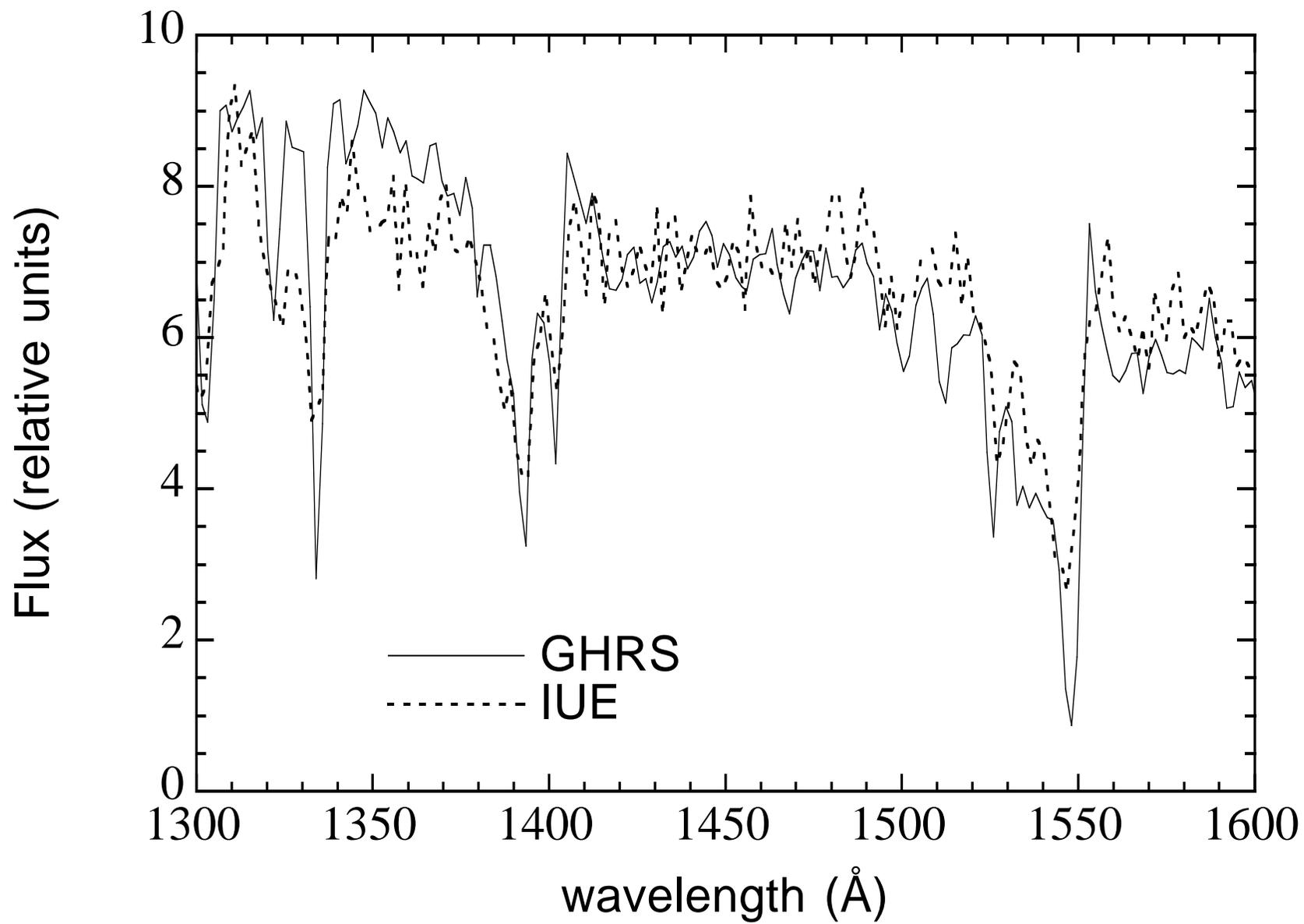
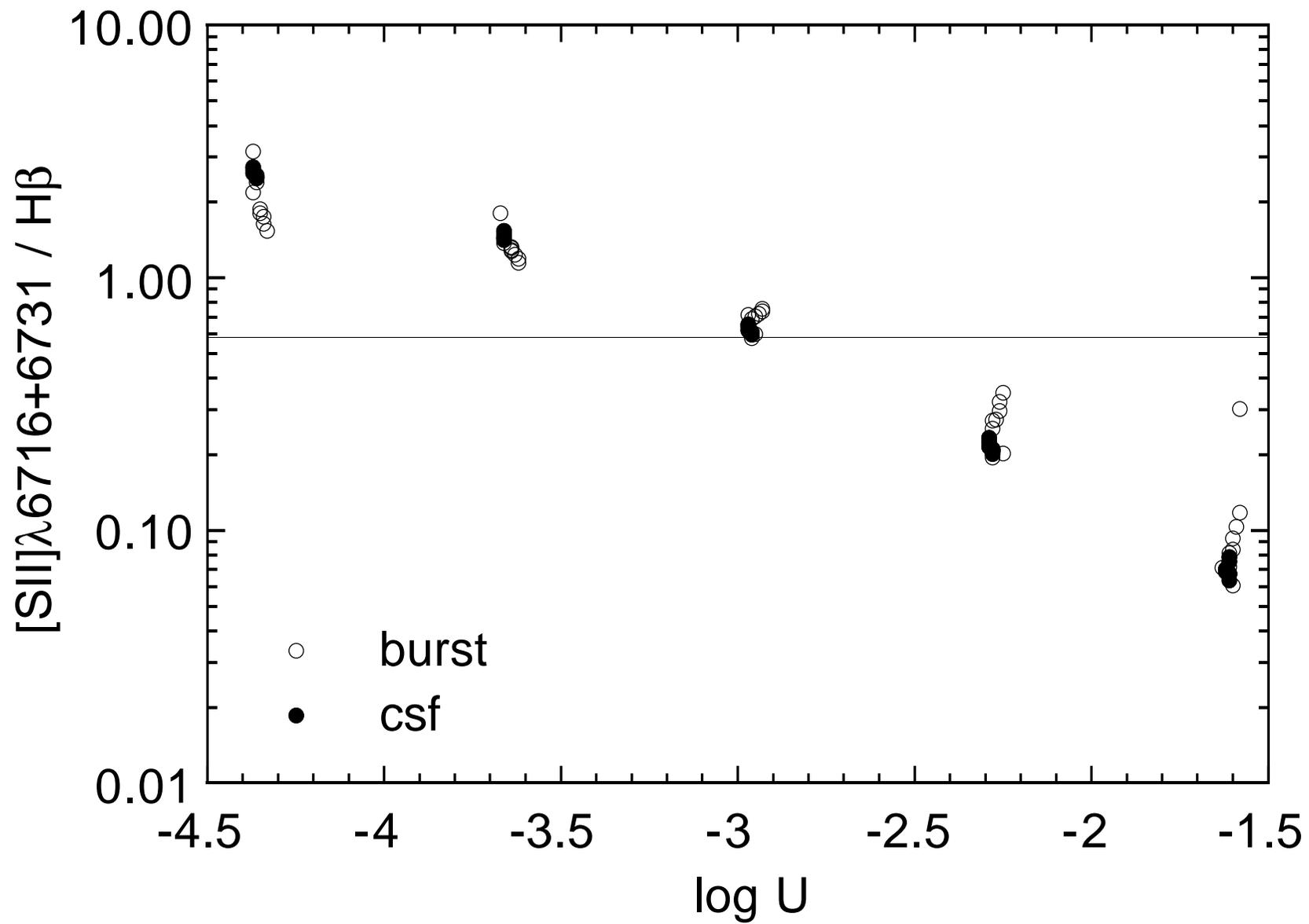
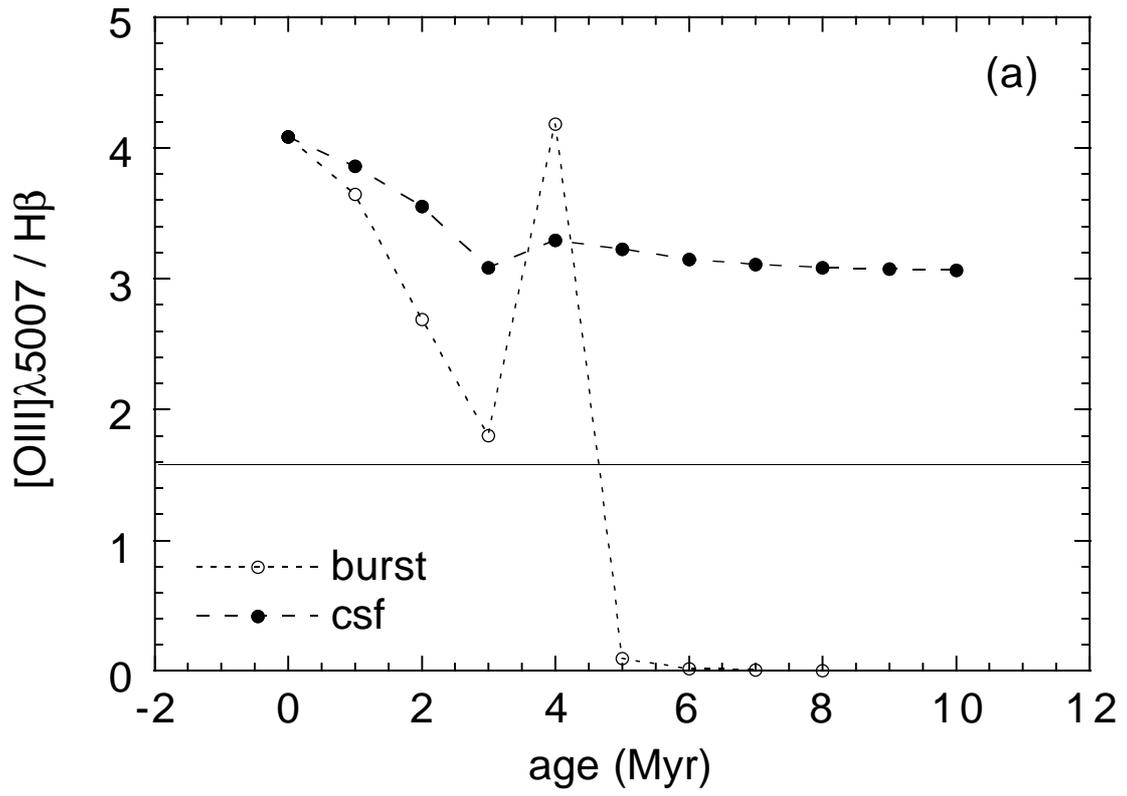
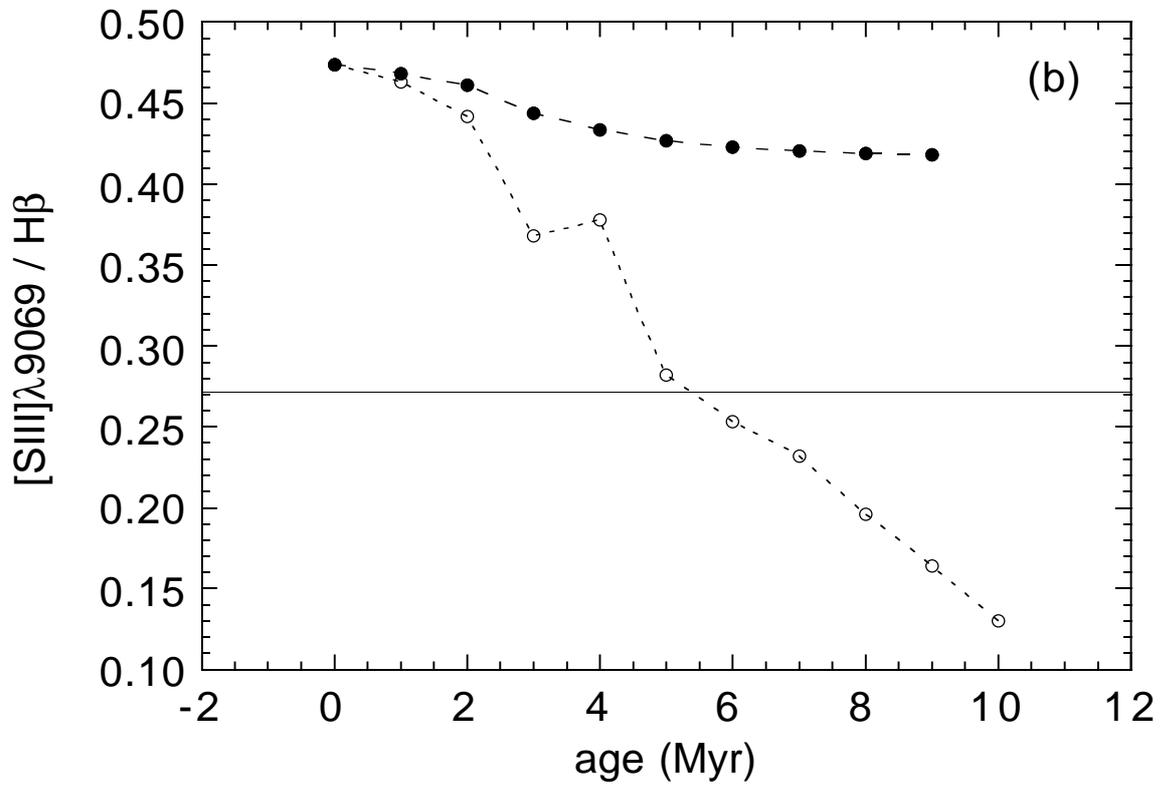
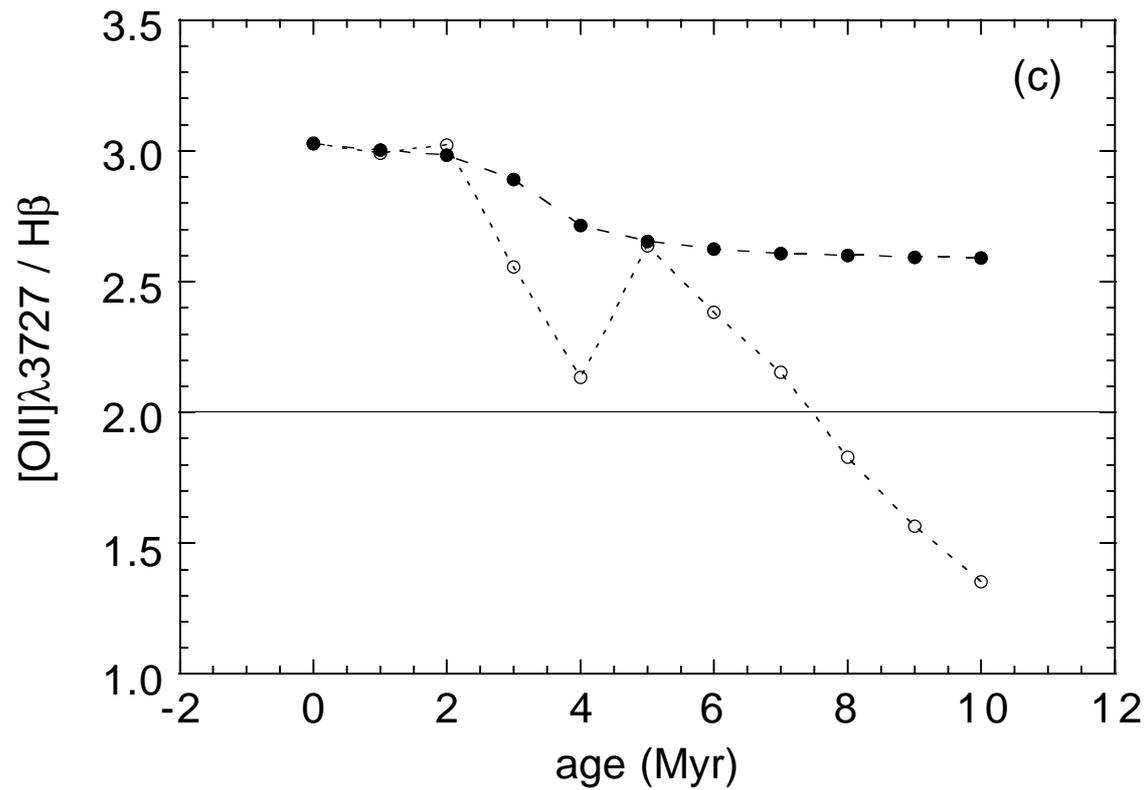
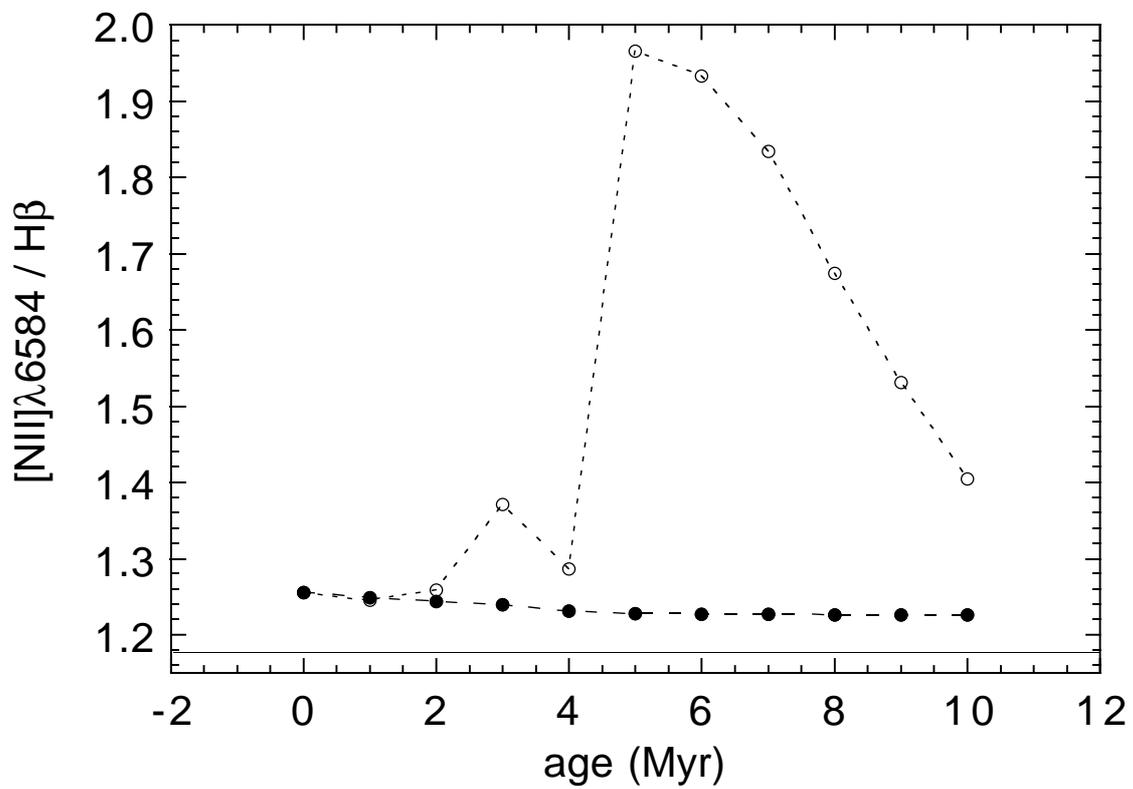
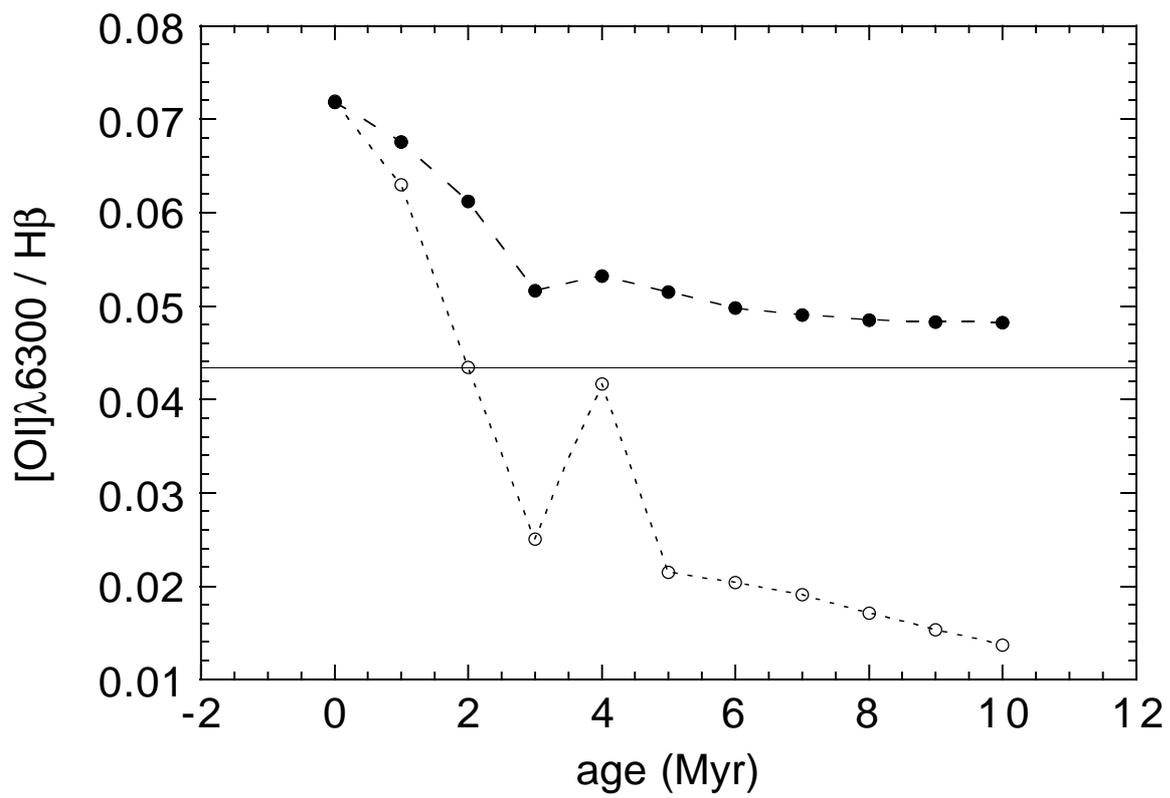
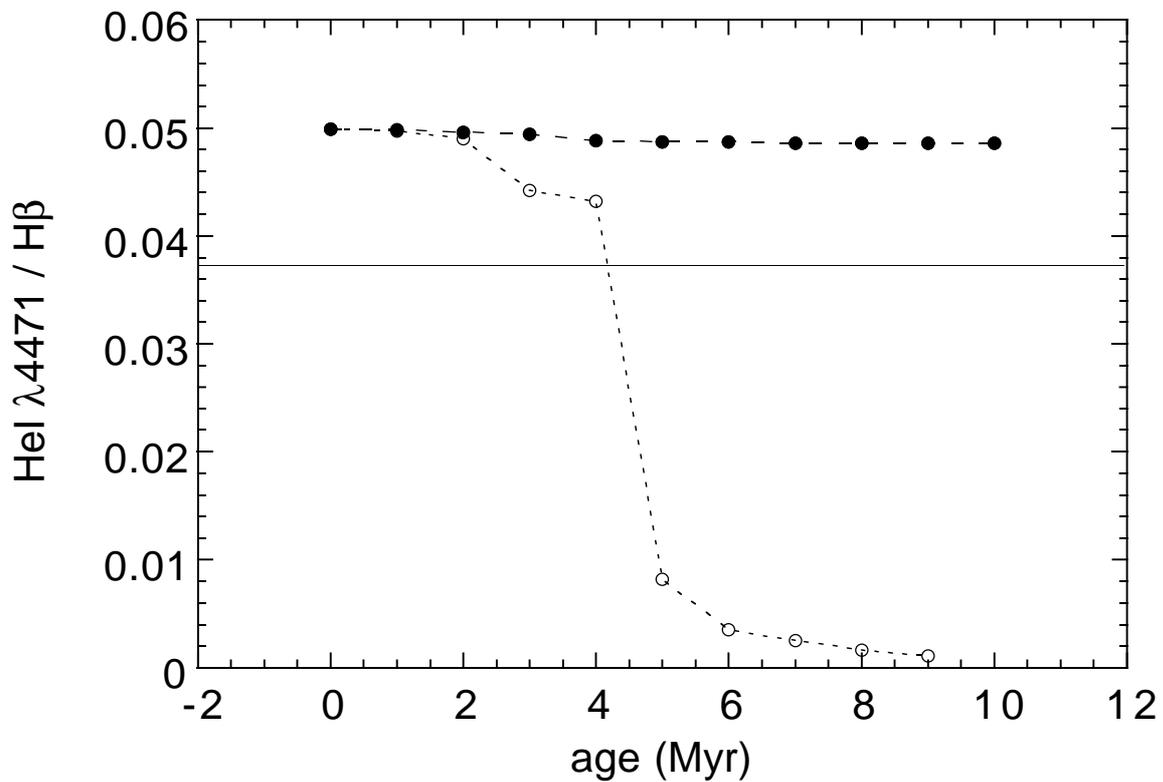
\end{document}